\documentclass[11pt,letterpaper]{article}
\pdfoutput=1
\usepackage{float}
\usepackage{verbatim}
\usepackage{graphicx}
\usepackage{feynmp}
\usepackage{slashed}
\usepackage{subfig}
\usepackage{jcappub}

\newcommand{\nco}{\newcommand}
\nco{\beq}{\begin{equation}} \nco{\eeq}{\end{equation}}
\nco{\beqa}{\begin{eqnarray}} \nco{\eeqa}{\end{eqnarray}}

\renewcommand{\b}{\bar}
\renewcommand{\comment}[1]{}

\newcommand{\X}{{\scriptscriptstyle X}}
\newcommand{\N}{{\scriptscriptstyle N}}

\title{Nonabelian dark matter models for 3.5 keV X-rays}

\author[a]{James M.~Cline}
\author[b]{and Andrew R.~Frey}

\affiliation[a]{Department of Physics, McGill University, 3600 rue University,
Montr\'eal, Qu\'ebec, Canada H3A 2T8}
\affiliation[b]{Department of Physics and Winnipeg Institute for Theoretical 
Physics, University of Winnipeg, Winnipeg, Manitoba, Canada R3B 2E9}

\emailAdd{jcline@physics.mcgill.ca}
\emailAdd{a.frey@uwinnipeg.ca}

\abstract{
A recent analysis of \textit{XXM-Newton} data reveals the possible
presence of an X-ray line at approximately 3.55 keV, which is not 
readily explained by known atomic transitions.  Numerous models of
eV-scale decaying dark matter have been proposed to explain this
signal.  Here we explore  models of multicomponent nonabelian dark
matter with typical mass $\sim$ 1-10 GeV  (higher values being allowed
in some models) and  eV-scale splittings that arise naturally from the
breaking of the nonabelian gauge symmetry.   Kinetic mixing between
the photon and the hidden sector  gauge bosons can occur through a
dimension-5 or 6 operator.   Radiative decays of the excited states
proceed through transition magnetic moments that appear at one loop. 
The decaying excited states can either be primordial or else produced
by upscattering of the lighter dark matter states.  These models are
significantly constrained by direct dark matter searches or cosmic
microwave background distortions, and are potentially testable in
fixed target experiments that search for hidden photons.  We note that
the upscattering mechanism could be distinguished from decays in
future observations if sources with different dark matter velocity
dispersions seem to require different values of the scattering cross
section to match the  observed line strengths.}

\keywords{dark matter theory}
\arxivnumber{}

\begin{document}
\maketitle

\section{Introduction}\label{s:intro}

Aside from its total mass density, little is known about
the particle nature of dark matter (DM).  Only upper limits exist on
its possible nongravitational
interactions with the Standard Model (SM), or on its self-interactions
that could come from dynamics of a hypothetical dark sector extending
beyond the dark matter itself.
Hints of positive detection from a variety of direct searches
\cite{Bernabei:2008yi,Angloher:2011uu,Aalseth:2011wp,
Aalseth:2010vx,Aalseth:2014eft,Agnese:2013rvf} are in apparent
conflict with limits from other experiments 
\cite{Angle:2011th,Aprile:2012vw,Aprile:2012nq,Akerib:2013tjd,Agnese:2013jaa},
presenting an increasingly difficult challenge for theorists to find nonminimal
models that could accommodate both kinds of results.
Anomalies in astrophysical observations have also been interpreted as harbingers
of the interaction of dark matter with the visible sector.
These include observations of excess positrons in the 10 GeV-TeV range
\cite{Adriani:2008zr,Barwick:1997ig,Chang:2008aa,Torii:2008xu}, 
a narrow feature in gamma rays at 130 GeV 
\cite{Bringmann:2012vr,Weniger:2012tx}, a gamma ray excess at energies
below 10 GeV
\cite{Hooper:2011ti,Abazajian:2012pn,Hooper:2010mq,Goodenough:2009gk}, and the long-studied galactic bulge positron 
population \cite{jhh,lms}, among others, as candidate signals
of DM.  Whether any of these observations ultimately prove to be related to
DM, they have led to a greater understanding of the range of physics 
possible in the dark sector.

Recently, refs.\ \cite{Bulbul:2014sua,Boyarsky:2014jta} identified an X-ray 
line with energy of approximately 3.55 keV in \textit{XMM-Newton} 
observations of galaxy clusters and the M31 galaxy, that is not associated
with any known atomic transition that could be consistent with the
observed intensity.  The line also 
appears in \textit{Chandra} observations of the Perseus cluster
\cite{Bulbul:2014sua}.  In the absence of a clear astrophysical explanation,
the possibility that this line is associated with DM is 
tantalizing; as proposed by the original two references, the decay of a
sterile neutrino to a photon and SM neutrino is a well-motivated DM 
explanation (see also \cite{Ishida:2014dlp,Abazajian:2014gza,Baek:2014qwa,
Shuve:2014doa,Tsuyuki:2014aia,Bezrukov:2014nza,Modak:2014vva,Robinson:2014bma,Chakraborty:2014tma,Haba:2014taa}).  
Alternative light DM candidates suggested as
the source of the X-ray line include axions and axion-like
particles \cite{Higaki:2014zua,Jaeckel:2014qea,Lee:2014xua,Cicoli:2014bfa,
Dias:2014osa}, 
axinos \cite{Kong:2014gea,Choi:2014tva,Liew:2014gia},
moduli \cite{Krall:2014dba,Nakayama:2014ova,Nakayama:2014cza}, 
light superpartners
\cite{Kolda:2014ppa,Bomark:2014yja,Demidov:2014hka,Dutta:2014saa}, 
and others \cite{Queiroz:2014yna,Dudas:2014ixa,Babu:2014pxa}.  
It is also possible that more massive (GeV-scale or higher) 
multi-species DM with a transition dipole moment
or other higher-dimension coupling can generate the 3.5-keV X-ray line 
\cite{Finkbeiner:2014sja,Aisati:2014nda,Frandsen:2014lfa,Allahverdi:2014dqa,Cline:2014eaa,Okada:2014zea,Lee:2014koa,Dutta:2014saa}.

While recognizing that the slow decay of a relic excited DM  state could account
for the X-ray line, ref.\ \cite{Finkbeiner:2014sja} pointed out  that collisional
excitation of DM followed by a relatively rapid decay could also do so.  This
mechanism of exciting dark matter (XDM) leads to a  distinct emission morphology
(following the dark matter density profile squared rather than linearly) and was
considered previously to address the galactic 511 keV emission (see
\cite{Finkbeiner:2007kk,Pospelov:2007xh,Finkbeiner:2008gw,ArkaniHamed:2008qn,Chen:2009dm,Finkbeiner:2009mi,Batell:2009vb,Chen:2009ab,Chen:2009av,Cline:2010kv,Vincent:2012an,Cline:2012yx,Bai:2012yq,Frey:2013wh}).  
In this paper, we will explore these scenarios in detail in the context of 
spontaneously broken nonabelian DM models, that can naturally have the necessary 
ingredients of small mass splittings  \cite{Thomas:1998wy} and kinetic mixing
with the photon through a dimension-5 or 6 operator.

\begin{figure}[t]\begin{center}
\includegraphics[scale=0.3]{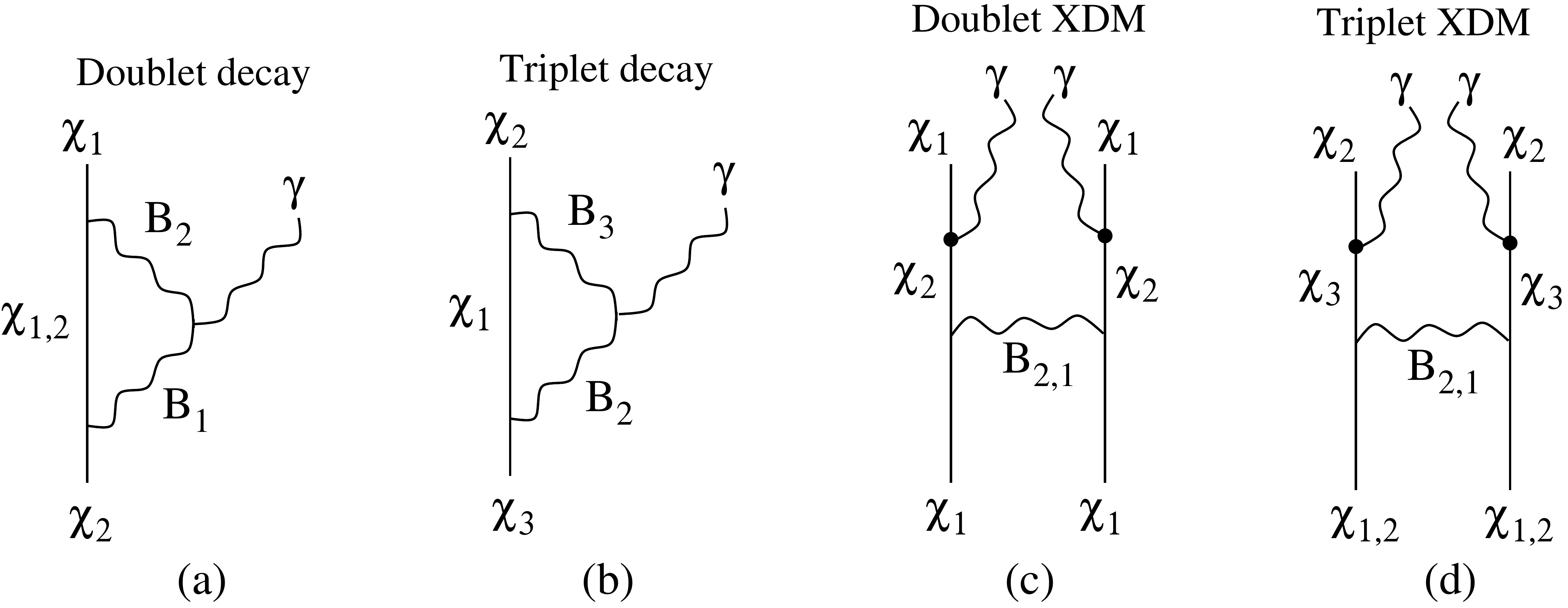}
\end{center}
\caption{(a,b) Diagrams for slow decay of relic excited state to
lower state in the doublet and triplet DM models, respectively;
(c,d) upscattering of lighter DM states to excited state, followed
by fast decays back to lower state (XDM mechanism), in respective models.} 
\label{fig:loop}
\end{figure}

We consider relatively simple hidden sectors, in which
the DM is a  Dirac or Majorana fermion 
$\chi_a$ transforming as 
a doublet or triplet respectively of a hidden-sector 
SU(2) gauge symmetry.  It is spontaneously broken by the vacuum
expectation value
(VEV) of a dark Higgs doublet
in the doublet DM model, or by either two dark Higgs triplets or a doublet
plus a triplet,  in the triplet DM model.
Significant kinetic mixing between the photon and one of the
components of the dark
gauge boson leads to a transition magnetic moment 
between DM states  \cite{Chen:2009ab}, by which the excited state
can decay to a lower state and a photon,
producing the 3.5 keV X-ray.\footnote{Ref.\ \cite{Cline:2010kv} 
predicted, long before the 3.5 keV line,
the existence of a several-keV X-ray line from exothermic models
of XDM, but
the expected value of flux was lower than required here, in the parameter
space of interest for explaining excess low-energy galactic positrons.}
The excited state 
might either be primordial in origin, with relatively long lifetime to explain 
the observed line, or it could be produced by
upscattering of the lower states, followed by fast decays.
The relevant processes are depicted in fig.\ \ref{fig:loop}.   

We begin in sect.\ \ref{s:signal} with a review of the observed X-ray
line and the general requirements for decaying or XDM
mechanisms to match the observed line strength.  Section
\ref{s:nonabelian} describes generic theoretical predictions of the
models that are relevant to our study,  including kinetic mixing,
relic density, mass splittings, magnetic moments, and cross sections
for inelastic self-scattering as well as scattering on protons.  In
section \ref{s:doublet} we confront these predictions to the
experimental constraints on models of doublet dark matter, while
section \ref{s:triplet} does likewise for triplet DM.
In section \ref{s:discussion} we summarize our findings 
and discuss the relation of nonabelian DM 
models to other current anomalies that may be indirect signals of 
dark matter.  Appendices give details of the computation of transition
magnetic moments and cosmic microwave background (CMB) constraints.

\section{XDM and the observed signal}\label{s:signal}

We begin by summarizing the requirements on the lifetime or
upscattering cross section from the observed line strength for the
decaying or XDM mechanisms, respectively.
For the decaying scenario, refs.\ \cite{Bulbul:2014sua,Boyarsky:2014jta} 
found that the 3.5 keV line could be produced by
sterile neutrino DM
of mass $m_s\sim 7$ keV with a lifetime of $\tau_s\sim 6.2\times 10^{27}$ s
(ref.\ \cite{Boyarsky:2014jta} gives errors of about a factor of 3 in either
direction).  Here we consider instead a heavy DM candidate with 
several nearly degenerate states, including a metastable one $\chi_x$
that decays to a lighter DM
state plus a photon.  Having in mind GeV-scale DM with a fractional abundance 
$f_x$ in the excited state, the required lifetime is
\beq 
	\tau = f_x \left(\frac{m_s}{M_\chi}\right)\tau_s = 
	(4.3\times 10^{21}\, \text{s})\,f_x \left(\frac{10\text{ GeV}}{M_\chi}\right)
	\implies
\frac\Gamma M_\chi = {1.5\times 10^{-47}\over
f_x}\ .
	\label{xray_rate} 
\eeq

As in refs.\ \cite{Finkbeiner:2014sja,Aisati:2014nda,Frandsen:2014lfa,Allahverdi:2014dqa,Cline:2014eaa,Okada:2014zea,Lee:2014koa}, 
we are interested in a decay $\chi_x\to\chi_g\gamma$ to the ground
state $\chi_g$ (or possibly to another long-lived excited state) 
plus a photon.  In our models this occurs 
via a transition dipole moment 
$\mu_{\times}\,\bar\chi_x\,\sigma_{\mu\nu}\chi_g\,F^{\mu\nu}$.
For mass splitting $\delta M_\chi$, the decay rate corresponding to
 this operator is
\beq 
	\Gamma = \frac{4\mu_{\times}^2}{\pi}\, \delta M_{\chi}^3\ ,
\eeq
so the required dipole moment for the X-ray signal is
\beq
	\mu_{\times}={1.7\times 10^{-15}\over\text{GeV}}
	\sqrt{\frac{M_\chi/f_x}{10\text{ GeV}}}\ .
	\label{reqdipole}
\eeq

\begin{figure}[t]\begin{center}
\includegraphics[scale=0.6]{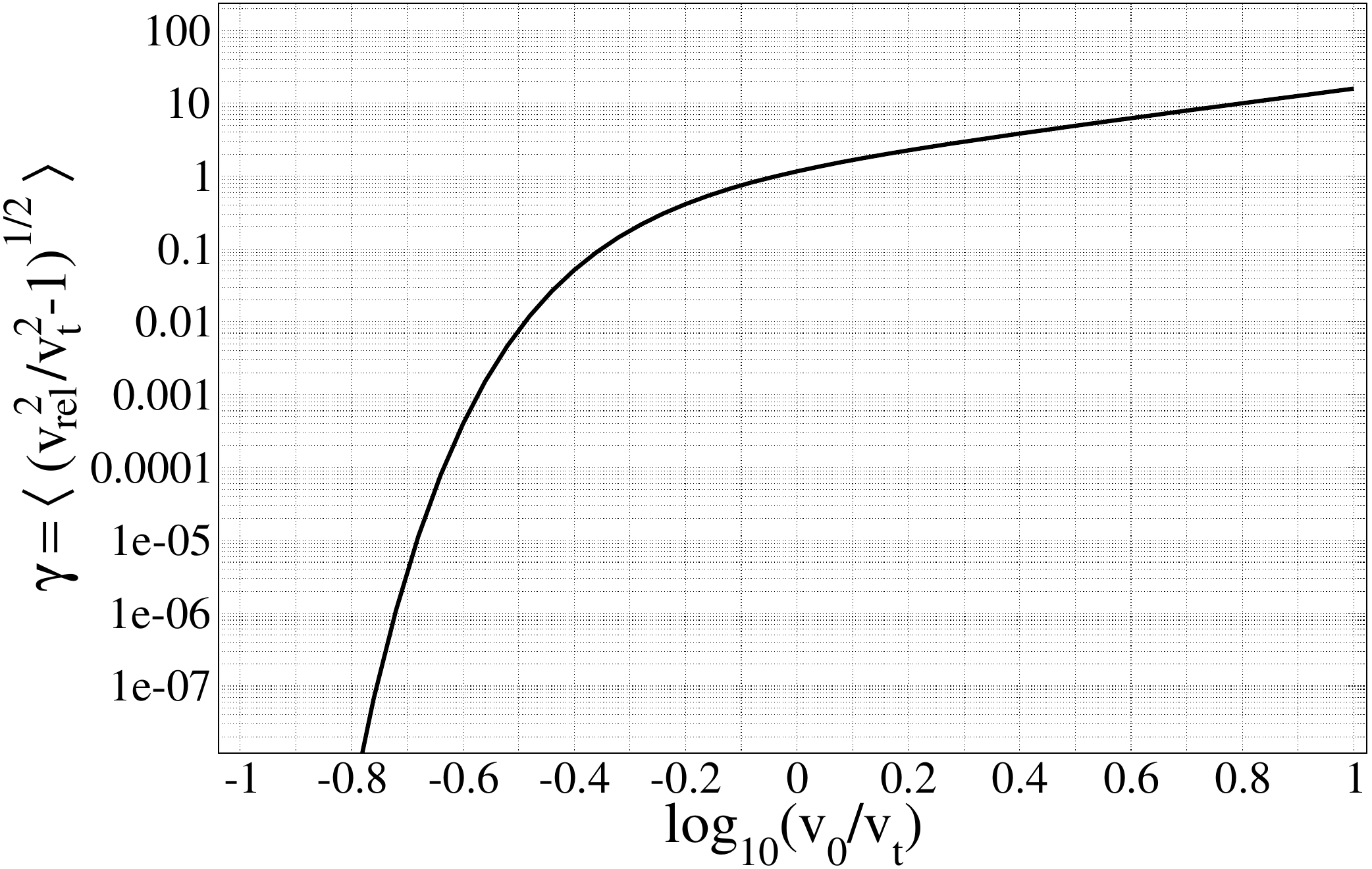}
\end{center}
\caption{Phase space average of $({v_{\rm rel}^2/v_t^2-1})^{1/2}$,
appearing in the upscattering cross section (\ref{sigma0}), for Maxwellian velocity
distribution $\sim\exp(-(v/v_0)^2)$, as a function of $v_0/v_t$,
where $v_t = \sqrt{8\delta M_\chi/M_\chi}$ is the threshold velocity.}
\label{f:thermavg} 
\end{figure}

On the other hand, if the transition dipole moment is larger than
(\ref{reqdipole}), the excited DM state will decay too rapidly, so there must
be some mechanism to repopulate it.  In XDM, this is accomplished through
collisional excitation.  In this case, the flux from a cluster or galaxy
at distance $d$ is
\beq
\label{xdmflux} 
	F = \frac{\eta_\X f_g^2}{4\pi d^{2} g_\chi} \int d^{\,3} x\, 
	\frac{\rho_\chi^2}{M_\chi^2} \langle\sigma_\uparrow v_{\rm rel}
	\rangle
\eeq
where $\sigma_\uparrow$ is the upscattering cross-section 
$\rho_\chi$ is the DM mass density; 
$f_{g}\sim 1$ is the fractional abundance of the DM ground state 
(or possibly a cosmologically long-lived
excited state), $g_\chi=2$ or $4$
depending on whether the DM is Majorana  or Dirac, and $\eta_\X$ is the number of
X-rays produced per collision ($\eta_\X=2$ in our models).  The cross section
is dominated by contributions near the kinematic threshold, so we approximate
\beqa
\label{sigma0}
	\langle\sigma_\uparrow v_{\rm rel}\rangle 
	&=& \sigma_0\,v_t\, \gamma; \nonumber\\
	\gamma &\equiv& \left\langle\sqrt{v_{\rm rel}^2/v_t^2-1}\,\,\Theta(v_{\rm rel}-v_t)
	\right\rangle,
\eeqa
where $v_t=\sqrt{8\,\delta m_{\chi}/M_\chi}$ 
is the threshold velocity for producing two excited states.
For the phase space average, we assume a Maxwellian distribution
$f(v) = N\exp(-(v/v_0)^2)$, where the velocity dispersion is
$\sigma_v = \langle v^2\rangle^{1/2} = \sqrt{3/2}\,v_0$.  Refs.\ 
\cite{Chapman:2006sg,Saglia:2009tp} find $\sigma_v=150$ km/s and 170 km/s
respectively for M31; we take the median value 160 km/s $= 5.3\times
10^{-4}\,c$.
The corresponding value for the Perseus cluster is 1300 km/s 
\cite{1980A&A....82..322D,1983AJ.....88..697K}.
Fig.\ \ref{f:thermavg} shows the dependence of
$\gamma$ on $v_0/v_t$.  If $v_0\gg v_t$ for both M31 and the Perseus
cluster, then the ratio of velocities $1300/160 = 8.1$ translates into
a similar ratio of fluxes for the two systems.  Below we will find 
a somewhat larger ratio $\gtrsim 20$.  While the quality of the
determinations is not sufficiently high to trust this number, if
correct, it could be accommodated by taking $v_0\cong 0.65\, v_t$
for M31, fixing $v_t \cong 6.7\times 10^{-4}\,c$.  With 
$v_0\cong 0.3\, v_t$ in M31, we can explain a factor of 100 difference
between the cross sections for the two systems.

To estimate the required cross section in (\ref{sigma0}), we 
compare to the observations of the 
X-ray flux for M31 and the Perseus cluster.
The DM halo of M31 can be modeled by an Einasto profile
\beq 
	\rho(r) = \rho_{-2}\, \exp\left[-\frac 2\alpha\left(\left(r\over
	r_{-2}\right)^\alpha-1\right)\right]
	\eeq
Ref.\ \cite{Tamm:2012hw} finds two fits 
with $\alpha=1/6$, normalization $\rho_{-2}=(8.9$ or $1.5)\times 10^{-2}$ 
GeV/cm$^3$, and scale radius $r_{-2}=(17.44$ or $37.95)$ kpc 
respectively.  
The field of view for the on-center observations of
M31 reported in \cite{Boyarsky:2014jta} is approximately 480 square
arcmin, corresponding to a radius of 2.8 kpc at the distance $d=785$ kpc,
and the flux is $F\approx 5\times 10^{-6}\,$s$^{-1}$cm$^{-2}$.  
Computing the volume integral in (\ref{xdmflux}) using these two
profiles, we estimate 
\beq 
	{\eta_\X f_{g}^2\,\langle\sigma v_{\rm rel}\rangle\over g_\chi M_\chi^2} 
	\approx \left(5\times 10^{-24}\ \textnormal{to}\ 
	3\times 10^{-23}\right)\ 
	\textnormal{cm}^3\,\textnormal{s}^{-1}\,\textnormal{GeV}^{-2}\ .
	\label{andromedasigma}
\eeq
(This is far below the limit $\langle\sigma v_{\rm rel}\rangle \lesssim 1$
b/GeV on the self-interactions of dark matter  from observations
of systems like the Bullet Cluster; see ref.\ 
\cite{Weinberg:2013aya} for a recent review.)

The analogous computation for the Perseus cluster gives
a volume integral of 
\beq 
	\frac{1}{4\pi d^2} \int d^{\,3} x\,\rho^2\approx 10^{16.25}
	\ \textnormal{GeV}^2\,\textnormal{cm}^{-5}
\eeq
for the entire cluster \cite{SanchezConde:2011ap}.
Ref.\ \cite{2010ApJ...710..634A} finds the similar value
$10^{16.15}$.  The 
central value of flux from
Perseus was measured to be $5.2\times 10^{-5}\,$cm$^{-2}\,$s$^{-1}$ 
(MOS, \cite{Bulbul:2014sua}), $7.0\times 10^{-6}\,$cm$^{-2}\,$s$^{-1}$ (MOS, 
\cite{Boyarsky:2014jta}), or $9.2\times 10^{-6}\,$cm$^{-2}\,$s$^{-1}$ (PN, 
\cite{Boyarsky:2014jta}).\footnote{MOS and PN refer to the two different types of
CCD cameras on {\it XMM-Newton}, metal oxide semiconductor and pn-junction
respectively.}
These yield 
\beq 
{\eta_\X f_{g}^2\,\langle\sigma v_{\rm rel}\rangle\over g_\chi M_\chi^2}  
	\approx (5\times 10^{-22}\ \textnormal{to}\ 
4\times 10^{-21})\ \textnormal{cm}^3\,
\textnormal{s}^{-1}\,\textnormal{GeV}^{-2}\label{perseussigma}\ .
\eeq
The cluster values are systematically higher than those of M31 by a
factor of $\sim 100$.  However the ranges in (\ref{andromedasigma})
and (\ref{perseussigma}) are not necessarily correlated, so the 
actual ratio could smaller.  In particular, if the 
true ratio is $\sim 8$, this would be consistent with $v_t\ll v_0$
in fig.\ \ref{f:thermavg}, both for $v_0$ of M31 and of the Perseus cluster.
In summary, the required cross sections are of order $f_{g}^2\,\langle\sigma_0\,v_{\rm rel}\rangle/M_\chi^2\sim 
10^{-22}$ cm$^3\,$s$^{-1}\,$GeV$^{-2}$ for M31, and $\sim 10-100$
times larger for Perseus, with the difference possibly being
attributable to the higher DM velocities in the cluster.

\section{General features of nonabelian DM models}
\label{s:nonabelian}

Before investigating specific models with respect to the X-ray line,
we summarize general features of the nonabelian DM models.  We 
elaborate on results of ref.\ \cite{Chen:2009ab} concerning the 
kinetic mixing, mass
splittings, magnetic moment, thermal relic density, self-scattering
cross section, and interaction with nucleons here.  The dark sector
consists of a
fermionic DM multiplet $\chi$ of mass $M_\chi$ transforming under a 
nonabelian gauge
symmetry with vector bosons $B^a_\mu$ and coupling $g$. 
We will consider only the simplest case of SU(2) for our specific
examples, but in this section we give some results for general
SU(N).  
The symmetry is spontaneously broken by some combination of 
doublet/fundamental ($h_i$) or triplet/adjoint ($\Delta_a$) dark Higgs fields, 
which leads to at least one
component of $B_a$  kinetically mixing with SM hypercharge through
a dimension-5 or 6 operator,
\beq
	{1\over \Lambda}\Delta^a B_a^{\mu\nu}Y_{\mu\nu}
	{\rm\quad or\quad } {1\over \Lambda^2}
	(h^\dagger \tau^a h)\, B_a^{\mu\nu}Y_{\mu\nu}
\label{mixing}
\eeq  
where $\Lambda$ is some heavy scale.  Once the Higgs has gotten a VEV,
this will lead to kinetic mixing  of a particular component $\hat a$
of the vector with the photon,  We normalize it in the conventional
way as $-(\epsilon/2) B^{\hat a}_{\mu\nu} F_{\mu\nu}$. Diagonalizing
the photon-$B^{\hat a}$ kinetic terms gives $B^{\hat a}$ a coupling of
strength $\epsilon e$ to  the electromagnetic current, which is the
main portal between the dark and visible sectors.

{\bf Relic density.}  We will be generally be interested in scenarios where the dark matter has a bare
mass $M_\chi \gtrsim $ 1 GeV, while the gauge bosons have masses
$m_{B_a}$ that are smaller.  In the model where $\chi_i$ is a doublet
of SU(2), this bare mass term only exists if $\chi$ is Dirac, 
whereas a triplet fermion can be Majorana.  Therefore the doublet DM
model has a global charge and can be asymmetric, whereas the triplet
would be expected to get its relic density through thermal freeze-out.
In such a case, assuming that Yukawa couplings of $\chi$ to the
Higgs bosons are negligible, freeze-out is determined by $\chi\chi\to
BB$ through the gauge interactions, and one can constrain the coupling
strength $\alpha_g = g^2/4\pi$ via the relic density. 
Updating the results of ref.\ \cite{Cline:2010kv} 
in light of the more  accurate relic density cross section of
\cite{Steigman:2012nb}, we find
\beq
\alpha_g \approx 1.6\times 10^{-4}\left({M_\chi\over 10\ \text{GeV}}\right)
\label{relic_const}
\eeq
for the SU(2) triplet model, assuming $M_\chi\gtrsim 5$ GeV.  
According to ref.\ \cite{Chen:2009ab}, the doublet model
 requires $\alpha_g$ to be
$2.5$ times larger,\footnote{Ref.\ \cite{Cline:2010kv} did a more
careful computation of the triplet model relic density than 
\cite{Chen:2009ab}, so we take the value of $\alpha_g$ from the former
reference, rescaled by the group theory factors found in the latter
for the case of the doublet model.}
assuming only the symmetric
component contributes to the relic density, 
A larger gauge group or DM representation requires a smaller 
$\alpha_g$.  In the more general case where Yukawa couplings could
be responsible for the relic density by annihilation into light
Higgses, eq.\ (\ref{relic_const}) can 
be interpreted as an upper bound on $\alpha_g$ to avoid suppressing
the relic density.

{\bf Mass splittings.} 
Of particular importance to our discussion, the states of the DM multiplet
are generically split in mass due to symmetry breaking, 
both via Yukawa couplings
and by differing gauge boson masses entering the DM self-energies
\cite{Thomas:1998wy}.  For $m_{B_a}\ll M_\chi$, the latter effect gives
rise to a correction to the DM mass term,
\beq
	\label{massmatrix}
	M\delta_{ij} -\frac{\alpha_g}{2}\sum_{B_a} m_{B_a} 
	\left(T^a T^a\right)_{ij}\ 
\eeq
For DM in the doublet representation of SU(2), the square of 
any generator (since it is a Pauli matrix) is the unit matrix,
so no splitting arises from this mechanism, and we are forced to
rely upon splittings generated by VEVs of dark Higgs fields that
 couple
to $\chi$.  For other representations, 
if all gauge boson masses are the same, the correction is 
proportional to the quadratic Casimir times $\delta_{ij}$, which 
leaves the states degenerate,
but in general 
$\delta M_\chi\sim \alpha_g \,\delta m_{B}/2$, where $\delta m_{B}$ is 
the typical splitting between the gauge boson masses.  Assuming that
$\delta m_{B} \sim m_{B_a}$ and $\alpha_g$ is given by
(\ref{relic_const}) the desired $\delta M_\chi$ of 
3.5 keV requires gauge boson masses of order 40 MeV$\times (10$
GeV$/M_\chi)$.  However it should be kept in mind that Yukawa
couplings can allow for larger $m_{B_a}$, both by directly contributing
to the mass splittings, and by allowing for smaller $\alpha_g$
as discussed above.

\begin{figure}[t]\begin{center}
\includegraphics[scale=0.5]{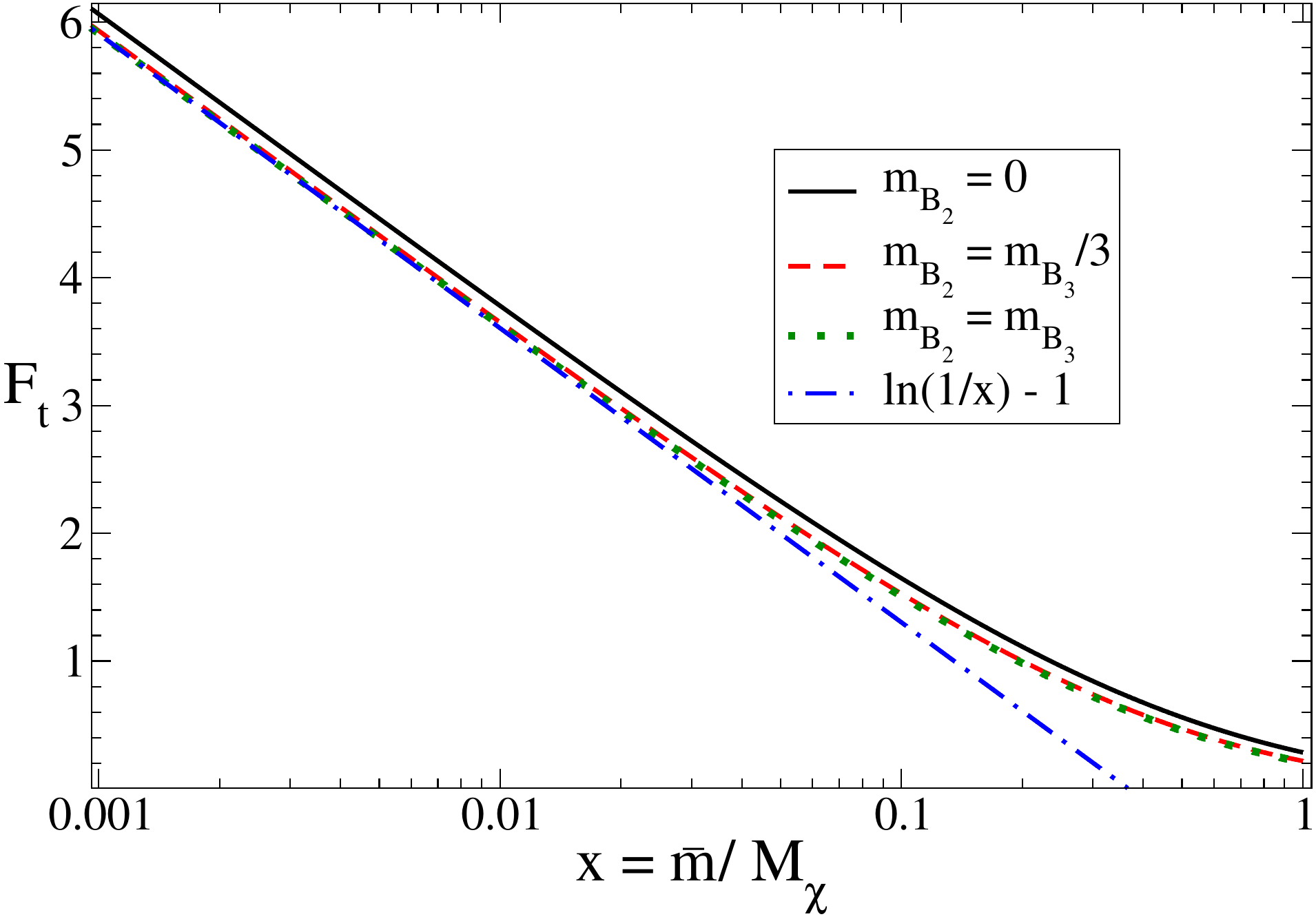}
\end{center}
\caption{\label{f:moment} 
The transition magnetic moment for triplet DM generated by 
kinetic mixing of a non-Abelian gauge group with the photon, $F_t(r_2,r_3)$
of (\ref{dipole}). $\b m^2 = (m_{B_2}^2+m_{B_3}^2)/2$, and the relation of $m_2$ to 
$m_{B_3}$ is labeled as in the legend for each curve.  The (blue) dot-dashed
curve is the $\b m\ll M_\chi$ approximation (\ref{dipole2}).} 
\end{figure}

{\bf Magnetic moments.} A unique feature of kinetic mixing in the form
(\ref{mixing}) is that it includes an interaction 
\beq
	\frac12 g\epsilon f^{\hat a bc} B^b_\mu B^c_\nu F^{\mu\nu}
\eeq
by virtue of the nonabelian field strength tensor $B^{\hat a}_{\mu\nu}$.
This operator generates transition (and in some cases direct) magnetic moments among the DM states,
as shown in figure \ref{fig:loop}(a,b) \cite{Chen:2009ab}.  
We calculate the transition moments 
in appendix \ref{app:mu}.  For models in which $\chi$ is in the
triplet representation of SU(2), we can take $\hat a = 1$ 
and find that the transition moment between
states 2 and 3 is given by\footnote{We
have corrected several factors of 2 relative to \cite{Chen:2009ab}.}
\beq
	\label{dipole} 
	\mu_{23} = \frac{\epsilon g^3}{16\pi^2\, M_\chi}
	F_t(r_2,r_3)
\eeq
where $r_i = (m_{B_i}/M_\chi)^2$.  The function $F_t$ is given in 
(\ref{ftfun}), but can be approximated as 
\beq
	F_t\sim \ln\left(\frac{M_\chi}{\b m}\right) -1\ 
	\label{dipole2}
\eeq
if $\bar m^2 \equiv (m_{B_2}^2+m_{B_3}^2)/2 \ll M_\chi^2$ and
if $|m_{B_2}-m_{B_3}|$ is not too large compared to 
$m_{B_1}$.  The behavior of $F_t$ is more generally illustrated in
fig.\ \ref{f:moment}, which shows that 
(\ref{dipole2}) is valid for 
$\b m\lesssim M_\chi/10$; above that value, 
$\mu_\times$ is
of order $\mathcal{O}(0.1-1) \epsilon g^3/16\pi^2 M$.
For models in which $\chi$ is an adjoint of SU(N), 
eq.\ (\ref{dipole}) is
generalized  by including the factor 
$(i/2) C_2(A) T^{\hat a}_{23}$ where $C_2(A)$ is the
quadratic Casimir of the adjoint representation.

In models of doublet DM with kinetic mixing of $B_3$, 
fig.\ \ref{fig:loop}(a), there is destructive
interference between the two diagrams where $\chi_1$ and $\chi_2$
are in the loop, leading to a suppressed transition moment
\beq
	\mu_{12} = {\epsilon g^3\,\delta M_\chi \over
	16\pi^2\, M_\chi^2}F_d(r)
\label{dipole3}
\eeq
where we ignore the gauge boson mass splittings (which anyway
vanish if SU(2) is broken only by a Higgs doublet)
and take $r=(m_B/M_\chi)^2$.   The function $F_d$ is given in
(\ref{fdfun}) and behaves as $1/2r$ for small $r$, showing that the
suppression is less severe than at first sight: $\mu_{12}\sim
\delta M_\chi/m_B^2$ rather than $\delta M_\chi/M_\chi^2$.

{\bf Inelastic self-scattering.}  The excitation of lower to higher
DM states by inelastic scattering, as shown in fig.\ 
\ref{fig:loop}(d) for the triplet model, has the cross section 
\beq 
	\langle\sigma_\uparrow v_{\rm rel}\rangle = 
	2\pi\alpha_g^2\, v_t\,\gamma\,\frac{M_\chi^2}{m_B^4}\mathcal{F}(v_{\rm rel}/v_t)
\label{upscatt}
\eeq
where $\gamma$ is as defined in (\ref{sigma0}) and $\mathcal{F}$ is a slowly varying function 
that goes to unity near threshold \cite{Cline:2010kv}.  
(Recall that $v_t = (8\,\delta M_\chi/M_\chi)^{1/2}$ is the 
threshold velocity for the inelastic process.)  

In the doublet model,  two components of the gauge bosons are
exchanged in the $t,u$-channels.  There is  a cancellation in
the $\chi_1\chi_1\to\chi_2\chi_2$ amplitude that is exact in the case
of degenerate gauge bosons, due to the group theory
factors $\sum_{a=1,2}(\tau^a_{12})^2 = 0$.  This changes
the cross section by  the replacement
\beq
	m_B^{-4}\to (m_{B_1}^{-2}-m_{B_2}^{-2})^2\equiv {(\delta
m_B^2)^2\over m_B^8}
\label{replacement}
\eeq
relative to (\ref{upscatt}).
In the following, we will assume that the XDM doublet is asymmetric
dark matter with such a gauge boson mass splitting, which requires
the VEV of a triplet Higgs in the dark sector, in addition to the
doublet Higgs, since $\delta m_B=0$ with only the latter.

{\bf Interaction with protons.}  Due to the kinetic mixing of
$B^{\hat a}$ with the photon, the component $B^{\hat a}$ couples
with strength $\epsilon e$ to protons, and thus mediates DM
interactions with nuclei.  (There is a similar but much smaller
contribution from the $Z$ boson that we ignore.)    The 
spin-independent cross section on protons is 
\beq
	\sigma_p = 16\pi\epsilon^2\, \alpha\,\alpha_g\, \frac{\mu_p^2}{m_B^4}
\label{sigma_p}
\eeq
where $\mu_p$ is the proton-$\chi$ reduced mass.  This will give
rise to stringent constraints from direct detection searches for some
of the scenarios we study in the remainder.

While the doublet DM model presented in section \ref{s:doublet} scatters
elastically from nuclei, the Majorana triplet DM models of section 
\ref{s:triplet} scatter inelastically (either endothermically or 
exothermically).  Since our focus is on astrophysical signals, we do not
carry out a detailed analysis of the effects of this inelasticity on 
the nuclear recoil spectrum but rather consider only the kinematic
scaling of the overall event rate as in \cite{Cline:2010kv}.  For endothermic
or exothermic scattering, we rescale $\sigma_p$ by
\beq\label{inelastic}
\left\langle (v^2\mp v_\delta^2)^{1/2}\,\Theta(v\mp v_\delta)
\right\rangle/\langle v\rangle,\qquad
v_\delta=\left(\frac{2|\delta M_\chi|}{\mu_{\N\chi}}\right)^{1/2}\eeq
respectively, where $\langle\cdots\rangle$ denotes the phase-space average in the 
standard halo model as described in \cite{DelNobile:2014eta}, $v$ is the
DM speed in the earth frame, 
$v_\delta$ is the threshold velocity for nuclear
scattering, and $\mu_{\N\chi}$ is the reduced mass for the
$\chi$-nucleus system.  We will find that LUX is the most constraining
experiment for the inelastic models, and will therefore take
xenon as the relevant nucleus for determining 
$\mu_{\N\chi}$.

\section{Doublet DM model}
\label{s:doublet}

In the simplest version of doublet DM, $\chi$ is a Dirac field that
has bare mass term $M_\chi\bar\chi^i\chi_i$.  
However, as noted above, the radiative correction to the mass of the doublet states
is proportional to the identity matrix and leads to no mass splitting.
There are two ways to remedy this situation.  (1)
Introduce a heavy SU(2)-singlet fermion $\psi$ with mass $M_0$ (that we
take to be Dirac) and with couplings $y \bar\chi h\psi+ {\rm h.c.}$.
If $h$ gets the VEV $(v,0)^T$ in just the upper component (as we can
choose with no loss of generality), then $\chi_1$ and $\chi_2$ get
a mass splitting by the see-saw mechanism of $\delta M_\chi =
(yv)^2/M_0$.  (2) Enlarge the dark gauge group to SU(2)$\times$U(1) and
break it to U(1) by the Higgs doublet, just as in the standard model.
This splits the gauge boson masses analogously to the $W$ and $Z$.
Moreover, simple kinetic mixing of the dark U(1) with the SM
hypercharge would lead to the same structure of couplings as we
outlined previously.  In addition, the dark $W$ bosons would be
millicharged under electromagnetism.  
Although option (2) may be interesting, we will confine our attention
to the first in this work, which is simpler in having no extra 
long-range forces to consider.  

\subsection{Long-lived decaying DM}

 We start with the scenario where 
$\chi_2$ is cosmologically long-lived and decays into $\chi_1 +
\gamma$.  By combining eq.\ (\ref{reqdipole}) with (\ref{dipole3}),
we find the constraint on the kinetic mixing parameter
\beq
	\epsilon = {1.5\times 10^{-10}\over \alpha_g^{3/2}\, \hat
	F_d(r)}\left(0.5\over f_x\right)^{1/2}\left(M_\chi\over 10{\rm\
	GeV}\right)^{1/2}\left(m_B\over 100{\rm\ MeV}\right)^2
\label{eps_doublet}
\eeq
where $r=(m_B/M_\chi)^2$ and $\hat F_d = 2r F_d$ such that $\hat F_d$
goes from 1 to $\sim 1.5$ for $r\in[0,1]$, and is approximated to better than
1\% by $\hat F_d \cong 1 + 0.716\, r^{1/2} - 0.248\, r$ in that
interval.

\begin{figure}[t]\begin{center}
\centerline{\includegraphics[scale=0.5]{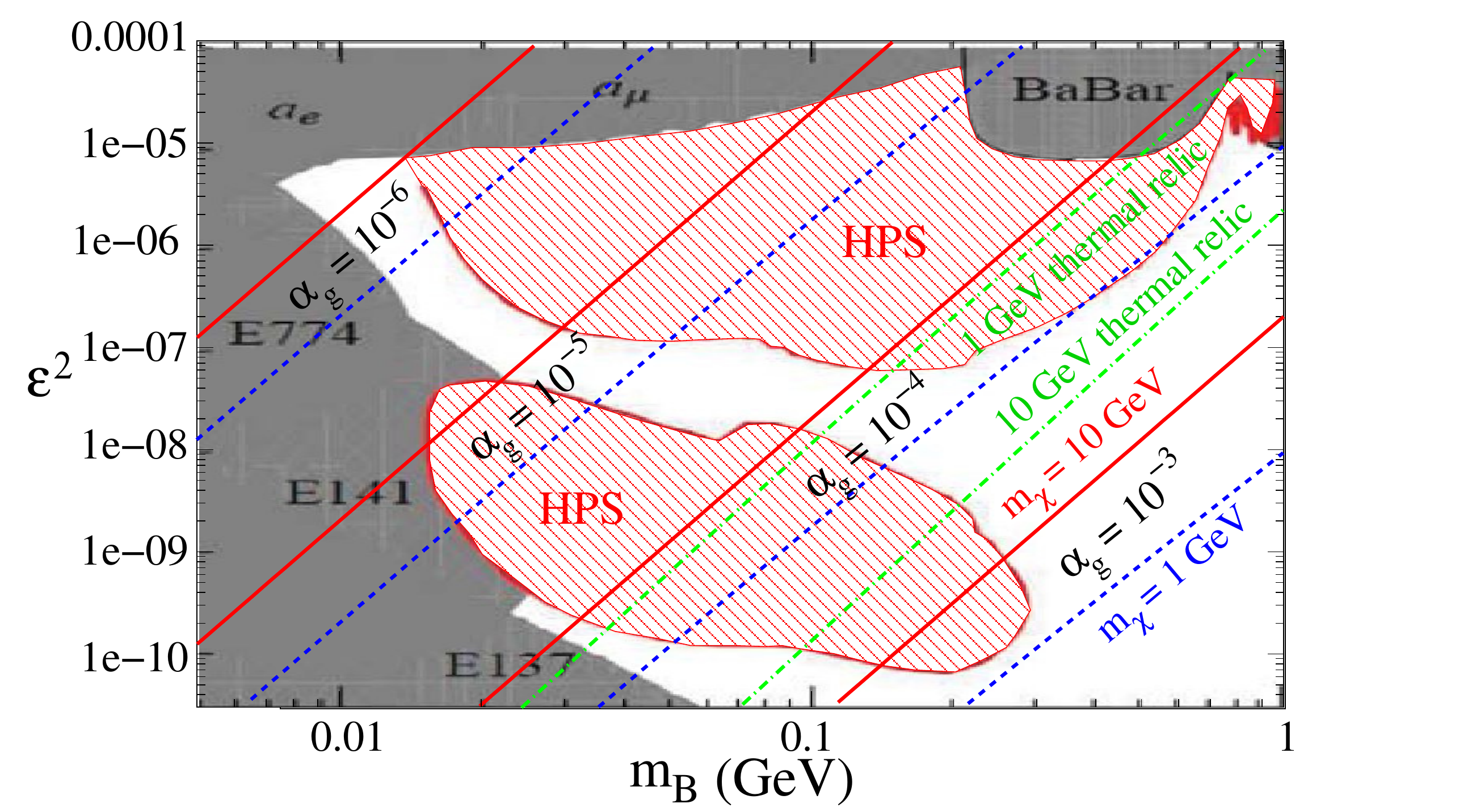}}
\end{center}
\caption{Dark shaded regions are excluded by searches for light
gauge bosons of mass $m_B$ coupling with strength $\epsilon e$ to electrons;
cross hatched region is to be probed by HPS experiment.
Diagonal lines are the values of $\epsilon$ needed for
doublet DM model to explain
the 3.5. keV X-ray line via decays of a long-lived excited state, assuming
indicated values of gauge coupling $\alpha_g$ and mass $M_\chi$. 
Dot-dashed lines assume value of $\alpha_g$ needed for thermal relic
density of $\chi$.  Background taken from ref.\ 
\cite{Stepanyan:2013tma}.} 
\label{f:hps-doublet}
\end{figure}

It is interesting to compare the prediction of (\ref{eps_doublet}) to
the sensitivity of existing and proposed searches for dark photons.
In fig.\ \ref{f:hps-doublet} we show the targeted region of the HPS
(Heavy Photon Search) experiment \cite{Stepanyan:2013tma}
in the $m_B$-$\epsilon$ plane,
and the constraint (\ref{eps_doublet}), 
assuming different values for $\alpha_g$ and $M_\chi$, and the
relative abundance $f_x=0.5$ for the excited state.
Regions of constant $\alpha_g$ are bounded by the solid and dashed
lines, corresponding to $M_\chi=10$ and $1$ GeV, respectively.
We note that there is intersection with the cross-hatched 
HPS regions of interest for a wide range of gauge couplings. 
Although the doublet model can be asymmetric and thus free from the
constraints associated with a thermal origin, 
we also show as dot-dashed lines the contours of $\alpha_g = 4\times
10^{-4}(M_\chi/10{\rm\ GeV})$ as needed for the thermal relic density,
for $M_\chi=10$ and $1$ GeV.  These also have significant overlap 
with the HPS regions, making this model more testable than many others
that have been proposed to explain the 3.5 keV X-ray line.  Gauge
couplings lower than the relic density bound need not be excluded even
for asymmetric DM, since there is another possible annihilation
channel $\chi\chi\to h h$ through the Yukawa interaction that splits
the $\chi$ masses.  This channel could be responsible for depleting
the symmetric component of $\chi$ for small values of $\alpha_g$. 

\begin{figure}[t]\begin{center}
\includegraphics[scale=0.5]{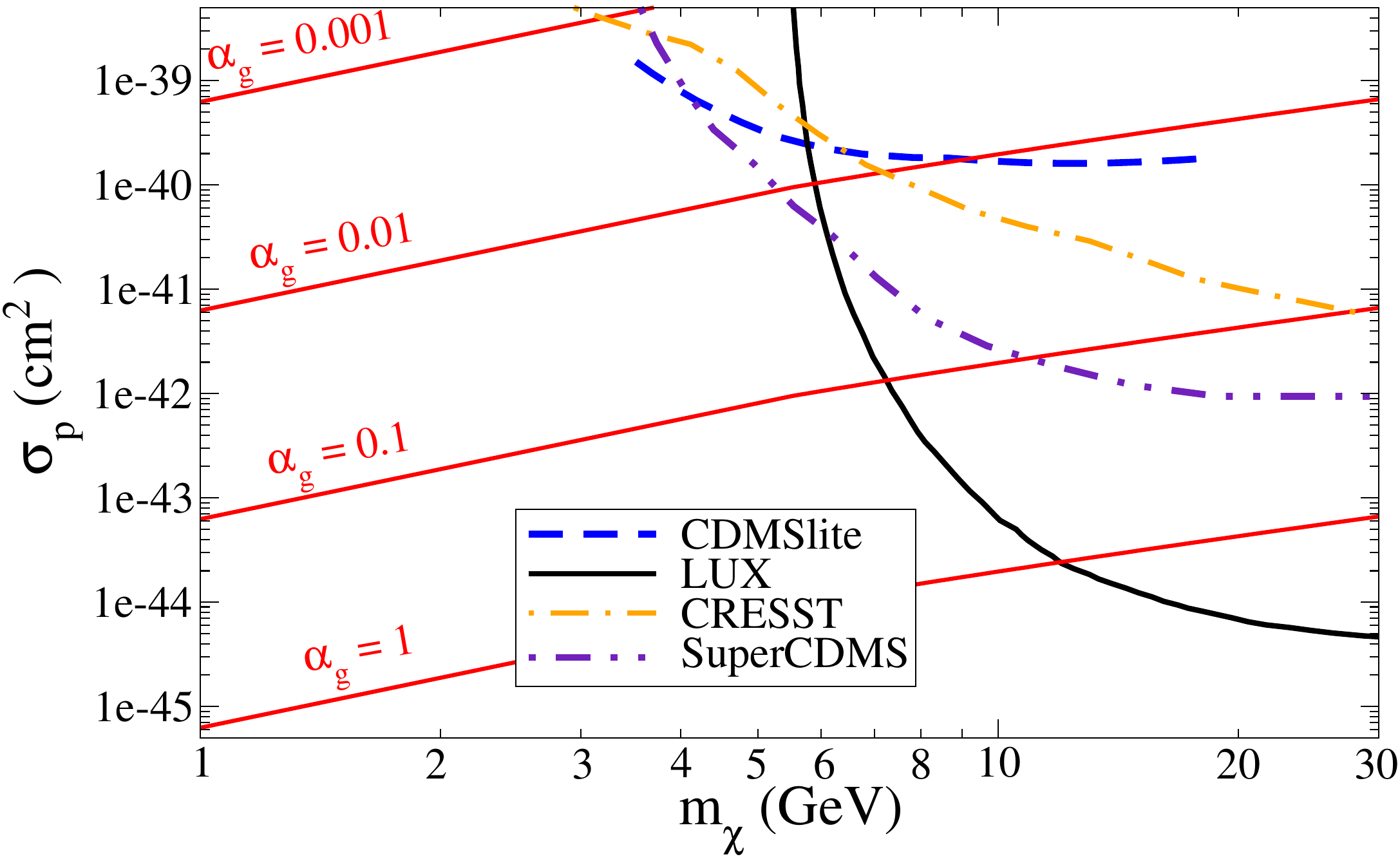} 
\end{center}
\caption{Cross section for doublet DM scattering on 
protons versus $M_\chi$, in the slowly decaying DM scenario,
along with current experimental limits.
Diagonal lines correspond to the model predictions for the
indicated values of 
gauge coupling, $\alpha_g = 10^{-3},\dots,1$, assuming 
$\epsilon$ is given by (\ref{eps_doublet}).}
\label{f:dd-doublet}
\end{figure}

We can also compare (\ref{eps_doublet}) to constraints from direct
detection searches, since the cross section for $\chi$ scattering on 
protons goes as $\epsilon^2$, eq.\ (\ref{sigma_p}).   Interestingly,
for $m_B\ll M_\chi$, the $m_B$-dependence cancels between
(\ref{eps_doublet}) and (\ref{sigma_p}) allowing for predictions
of $\sigma_p(M_\chi)$ that
depend upon only  one unknown parameter, $\alpha_g$.  Eliminating
$\epsilon$ leads to the cross section on protons
\beq
	\sigma_p = {4\times 10^{-45}\,{\rm cm}^2\over \alpha_g^2}
	\left(0.5\over f_x\right)
	\left(1 + {m_p\over M_\chi}\right)^{-2}\left(M_\chi\over 10{\rm\ GeV}\right)
\eeq
In fig.\ \ref{f:dd-doublet} we show the predicted value of $\sigma_p$
versus $M_\chi$ for a range of fixed $\alpha_g$, along with the
current upper limits from the LUX \cite{Akerib:2013tjd}, CDMSlite
\cite{Agnese:2013jaa}, SuperCDMS \cite{Anderson:2014fka}, and CRESST
\cite{Angloher:2014dua} experiments.  We have relaxed the 
published limits of 
the experiments by the factor $(Z/A)^2$ appropriate for each one, to
account for our model coupling only to protons and not all nucleons.
(For CRESST, it is assumed that the tungsten component
 gives the dominant constraint.)   It is clear that mainly models with
large values of $\alpha_g\gtrsim 10^{-3}$ that would pertain to asymmetric DM are
constrained by direct detection, and that $M_\chi$ must be rather
small, $\lesssim 10$ GeV, to escape detection.  Comparison with fig.\
\ref{f:hps-doublet} shows the complementarity of direct detection and
electron beam dump experiments for constraining the model.  Only in
the case $\alpha_g\sim 10^{-3}$, $M_\chi\sim 3$ GeV could there be
some overlap in coverage, allowing for discovery by both kinds of
experiments.

\begin{figure}[t]\begin{center}
\includegraphics[scale=0.5]{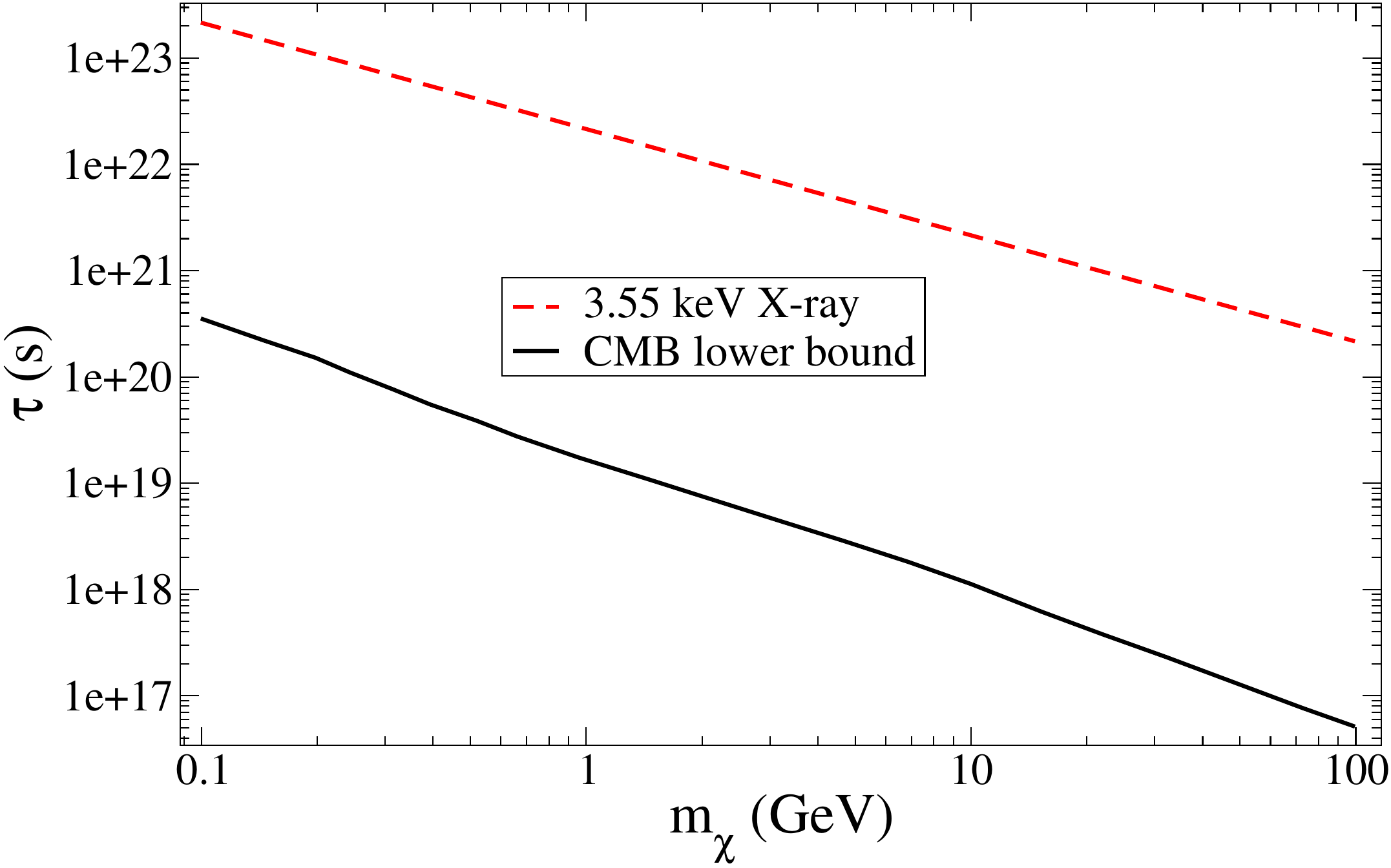} 
\end{center}
\caption{Solid curve: CMB lower bound on lifetime of excited state versus mass
for decays into 3.55 keV X-ray, adapted from ref.\
\cite{Diamanti:2013bia}.  Dashed: required value from 3.55 keV line
strength, eq.\ (\ref{xray_rate}). } 
\label{f:cmb-doublet}
\end{figure}

In terms of indirect limits, the strongest constraint on decaying DM
comes from distortions of the cosmic microwave background  due to
injection of electromagnetic energy at the time of recombination
\cite{Mapelli:2006ej,Zhang:2007zzh,Finkbeiner:2011dx,
Yeung:2012ya,Slatyer:2012yq,Cline:2013fm,Diamanti:2013bia}.  Usually
this is presented as a lower limit on the lifetime $\tau$ as a
function of mass $M_\chi$, assuming that all the mass-energy goes into
ionizing radiation.  If only a fraction $\delta M_\chi/M_\chi$ does so,
the constraint on the lifetime is loosened by this factor.  There
can be a compensating factor of ${\cal O}(1)$ for the greater efficiency
of absorption of keV energies relative to multi-GeV's 
\cite{Slatyer:2012yq}, but this does not affect our conclusions.
 We replot
the strongest constraint from ref.\ \cite{Diamanti:2013bia} in fig.\ 
\ref{f:cmb-doublet} taking account of this factor.  We have also 
applied an additional correction factor of $3.55$ (no relation to the
energy of the X-ray line) for the relative 
ionization efficiencies of photons and electrons \cite{Cline:2013fm}
that further weakens their limit.  The result is several orders of 
magnitude weaker than what is required to get the observed 3.55 keV
line strength, eq.\ (\ref{xray_rate}), also plotted in the figure.

\begin{figure}[t]\begin{center}
\centerline{\includegraphics[scale=0.5]{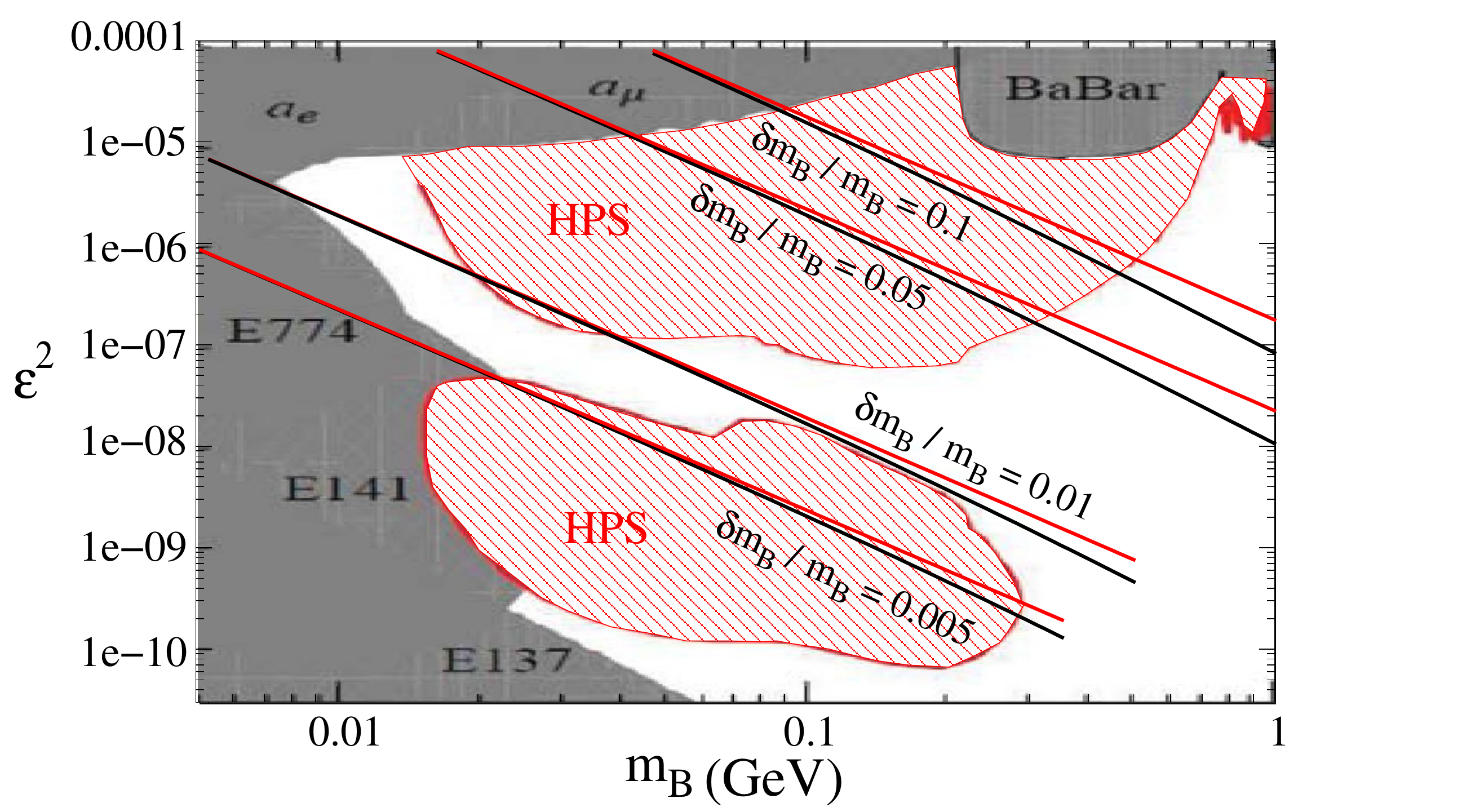}}
\end{center}
\caption{Like fig.\ \ref{f:hps-doublet}, but for the XDM version
of the doublet model.  Here the diagonal lines are {\it lower bounds}
from the requirement of sufficiently fast decay of the excited state.
$\alpha_g$ varies from $10^{-4}$ to $1$ as $m_B$ increases.
Upper (red) curves correspond to $M_\chi\gtrsim 10$ GeV, while
lower (black) are for $M_\chi = 1$ GeV.  Gauge boson mass splitting 
varies from $\delta m_B/m_B = 0.005$ to 0.1 as indicated.} 
\label{f:hps-xdm}
\end{figure}

\subsection{XDM doublet model}\label{s:doubletxdm}

As explained in section \ref{s:nonabelian}, it is necessary to introduce 
a mass splitting between the $B_{1,2}$ gauge bosons in order to
have nonvanishing inelastic scattering $\chi_1\chi_1\to \chi_2\chi_2$
for asymmetric doublet DM.\footnote{One might also consider symmetric
doublet DM, in which the inelastic scattering 
$\chi_1\bar\chi_1\to \chi_2\bar\chi_2$ can proceed through the
$s$ channel.  But it turns out that this gives too small a
cross section to explain the X-ray signal if $\alpha_g$ is as small as
required by the thermal relic density constraint (\ref{relic_const}).}
For example a triplet Higgs with VEV $\langle \Delta^a\rangle = 
\Delta\delta_{a,1}$ splits the gauge boson masses by 
$m_{B_2}^2 -  m_{B_1}^2 = 
m_{B_3}^2 - m_{B_1}^2 = g^2\Delta^2$ \cite{Chen:2009ab}.

In the scenario where $\chi_1\chi_1\to \chi_2\chi_2$ produces the
excited state, we can constrain $m_B$ by equating the theoretical
upscattering cross section (\ref{upscatt}, \ref{replacement}) to one of the
observational estimates in (\ref{andromedasigma}) or 
(\ref{perseussigma}).  For definiteness, taking the lower value
of (\ref{perseussigma}) we find
\beq
	m_B \cong \left(\delta m_B\over m_B\right)^{1/2}
	\left(\alpha_g^{1/2}\over 0.1\right)
\	{(v_t\gamma)^{1/4}\over 0.1} \times 280{\rm\
MeV}
\label{mBeq}
\eeq
where $\gamma$ is the quantity plotted in fig.\ \ref{f:thermavg}.

A further requirement in this
scenario is that the excited state must decay faster than the current
Hubble rate.  But in that case, 
CMB constraints are important, and in fact
require that the lifetime be
less than approximately $10^{12}\,$s, so that primordial contributions
have disappeared before the era of recombination (see for example
ref.\ \cite{Slatyer:2012yq}).  This puts a lower bound on the kinetic mixing:
\beq
	\epsilon > {1.1\times 10^{-6}\over \alpha_g^{3/2} 	
	\hat F_d(r)} \left(m_B\over 100{\rm\, MeV}\right)^2
	= {0.09\over \hat F_d(r)} 
	\left(\delta m_B\over m_B\right)
	\left(v_t\gamma\over \alpha_g\right)^{1/2}
\label{eps2xdm}
\eeq
where we used (\ref{mBeq}) to get the second expression.
To compare with the sensitivity region of the HPS experiment,
we vary $\alpha_g$ between $10^{-4}$ and $1$ to generate parametric
curves of $\epsilon$ versus $m_B$ from (\ref{mBeq}) and
(\ref{eps2xdm}), assuming $v_t \gamma = 10^{-3}$ for definiteness,
and considering $\delta m_B/m_B$ and $M_\chi$ for several discrete
values.  The result is 
shown in fig.\ \ref{f:hps-xdm}.  Since these curves only indicate
lower bounds on $\epsilon$, there is considerable overlap between
these predictions and the reach of HPS.

Moreover, (\ref{eps2xdm}) gives a lower bound for the cross section on
protons from (\ref{sigma_p}), 
\beq
	\sigma_p > {10^{-36}{\rm\, cm}^2\over \alpha_g^2\,\hat
	F_d^2}
	\left(1 + {m_p\over M_\chi}\right)^{-2}
\eeq
This
constrains $M_\chi < $ a few GeV in order to evade direct detection.
(Such small masses could be dangerous from the point of view of CMB 
constraints on annihilations resulting in electrons, except that we have 
assumed the
doublet is asymmetric dark matter in the XDM scenario.)
The XDM version of the model thus predicts an enormous signal for
future low-threshold direct detection experiments that would be
sensitive to light dark matter.

\section{Triplet DM model}\label{s:triplet}

Unlike doublet representation DM, in which both components couple 
equally to all gauge bosons, triplet DM components $\chi_a$ each couple only
to two of the three SU(2) gauge bosons $B^a$. As a result, radiative corrections
lead to mass splittings of order $\delta M_\chi\sim\alpha_g\,\delta m_{B}/2$
as in (\ref{massmatrix}).  Gauge boson mass differences of 
$\delta m_{B}\sim 45$ MeV$\times(10$ GeV$/M_\chi)$ therefore lead to the appropriate
splitting of $\delta M_\chi=3.55$ keV.  We will take $\chi_a$ to be Majorana
for simplicity.

\subsection{Mass Splittings}\label{s:tripletsplittings}
We consider two different dark Higgs sectors that turn out to have
interestingly different predictions for direct detection, due
to the spectra of gauge boson masses.
In the first case, there are two triplet Higgs fields, 
denoted $\Delta^a$ and $\Delta'^a$, 
with VEVs $\Delta^1=\Delta \delta^{a1}$ 
and $\Delta'^a=\Delta'\delta^{a2}$.  
For simplicity we take only one of $\Delta,\Delta'$ to appear in the
kinetic mixing operator (\ref{mixing}), so that only one gauge boson
mixes with the photon.\footnote{More generally, if the VEVs are not quite 
orthogonal, the
other gauge bosons pick up (smaller) kinetic mixings, as well, without
changing the conclusions below in a substantial way.}  In the
second case, there is a single triplet Higgs $\Delta^a$ with VEV $\Delta^a=\Delta\delta^{a2}$ 
and a doublet Higgs $h$ with VEV $(v/\sqrt 2)(1,1)^T$.  
Again for simplicity we assume that only one of these fields appears in the kinetic mixing 
(\ref{mixing}).\footnote{Throughout, we take $\Delta,\Delta',v>0$ without
loss of generality.}

In either case, symmetries forbid any Yukawa 
couplings, so the DM mass splittings come exclusively from 
radiative corrections
and for triplet DM take the form $\delta M_{ab}\equiv M_{\chi_b}-M_{\chi_a} = 
\alpha_g(m_{B_b}-m_{B_a})/2$.
With two triplet Higgs fields, the gauge boson masses are
\beq\label{gaugemasses1}
m_{B_1} = g\Delta', \quad m_{B_2} = g\Delta,\quad m_{B_3}=g\sqrt{\Delta^2+
\Delta'^2},
\eeq
giving rise to radiative DM mass splittings \cite{Chen:2009ab}
\beq\label{splittings1}
\delta M_{12} = \frac 12 g\alpha_g \left(\Delta -\Delta'\right),\qquad
\delta M_{23}=\frac 12 g\alpha_g \left(\sqrt{\Delta^2+\Delta'^2}-
\Delta\right)\ ,
\eeq
where $\chi_3$ is the heaviest DM state and has a transition
magnetic moment with either $\chi_1$ or $\chi_2$.  
With doublet and triplet
Higgs (whose VEV is in the $2$-direction), the gauge boson masses are
\beq\label{gaugemasses2}
m_{B_1}=m_{B_3} = g\sqrt{v^2+\Delta^2},\quad m_{B_2} =gv
\eeq
corresponding to DM mass splittings
\beq\label{splittings2}
\delta M_{21} =\delta M_{23} = \frac 12 g\alpha_g \left(
\sqrt{v^2+\Delta^2} - v\right).\eeq
$\chi_2$ is the lightest state, and $\chi_1,\chi_3$ are degenerate.  
We have chosen the $h$ VEV $= (v,v)^T/\sqrt{2}$ so that 
kinetic mixing of $B_1$ generates a transition
magnetic moment between $\chi_3$ and $\chi_2$.  

If all the Higgs VEVs are of the same order of
magnitude and the DM is produced thermally (so that $\alpha_g$
is determined by (\ref{relic_const})),
then $\{\Delta,\Delta',v\}\sim 1\text{ GeV}\times (10 \text{ GeV}/M_\chi)^{3/2}$
and therefore gauge boson masses of order $45$ MeV$\times(10\text{ GeV}/M_\chi)$ yield
the desired 3.55 keV mass splitting.  On the other hand, we can obtain 
$\delta M_{23}=$3.55 keV if $\Delta\gg \Delta'$ or $v\gg \Delta$ for the two
Higgs sectors respectively, leading to mass splittings
\beq 
	\delta M_{23} = \frac 14 g\alpha_g \left\{ 
	\frac{(\Delta')^2}{\Delta},
	\frac{\Delta^2}{v}\right\}\ ,
\eeq
This gives $(\Delta')^2=\Delta \times(2\text{ GeV})(10\text{ GeV}/M_\chi)$ 
in the case of two triplets, and 
gauge boson masses $m_{B_1}\gtrsim 400$ MeV and $m_{B_{2,3}}\gtrsim 2$ GeV,
with approximate equality at $\Delta\sim 5\Delta'$.  With one doublet and
one triplet Higgs, all the gauge boson masses are $m_{B_a}\gtrsim 2$ GeV.
Nonthermal production of DM requires a larger gauge coupling and therefore
allows for smaller Higgs VEVs and gauge boson masses.

\subsection{Decaying DM Model}\label{s:tripletdecays}

For triplet DM, the transition dipole moment $\mu_{\times}$ is
given by (\ref{dipole}).  If the X-ray line is produced through 
long-lived excited
state DM decay, the kinetic mixing parameter is 
\beq
\label{epstriplet}
	\epsilon = \frac{1.1\times 10^{-13}}{\alpha_g^{3/2} F_t(r_2,r_3)}
	\left(\frac{0.3}{f_3}\right)^{1/2}\left(\frac{M_\chi}{10\text{ GeV}}\right)^{3/2}
	\ ,
\eeq
where $r_{2,3}=(m_{B_{2,3}}/M_\chi)^2$.  Recall that the $B_1$ gauge
boson is the one that mixes with the photon, and its mass depends on the
dark Higgs sector of the model.  In models with symmetry broken either
by two triplet Higgs fields or by a triplet and a doublet, eqs.\
(\ref{gaugemasses1}-\ref{splittings2}) imply that 
$m_{B_3}-m_{B_2}=45$ MeV $\times(10$ GeV$/M_\chi) \equiv f(M_\chi)$;
however $m_{B_1}=m_{B_3}$ for doublet plus triplet Higgses whereas
$m_{B_1}^2 = m_{B_3}^2-m_{B_2}^2$ for two triplets.  This leads to
the gauge boson mass relations
\beqa
	m_{B_2} &=& m_{B_1}-f, \quad\quad\quad\ \  m_{B_3} = m_{B_1}, 
\	\quad\quad\quad\quad\quad{\rm doublet}
+ {\rm triplet\  Higgses}\nonumber\\
	m_{B_2} &=& \frac12(m_{B_1}^2/f - f), \quad m_{B_3}
	=\frac12(m_{B_1}^2/f + f),\quad{\rm two\ triplet\ Higgses}
\label{massrels}
\eeqa 
which fixes $m_{B_{2,3}}$ in terms of $m_{B_1}$ and $M_\chi$ in what
follows.

With any of the mass splittings and gauge boson masses discussed in
section \ref{s:tripletsplittings}, the fractional relic abundance of $\chi_3$
is $0.1\lesssim f_3\lesssim 0.33$ for thermal relic DM, as we have
verified using the methods described in 
\cite{Cline:2010kv}.  Since $F_t$ runs 
between approximately 1/2 and 4 in the parameter space of interest, we find
a required kinetic mixing $\epsilon\lesssim 10^{-8}$, which avoids 
current
laboratory constraints on light vector bosons (summarized in 
\cite{Morrissey:2014yma}).  However constraints on supernova cooling 
require $m_{B_1}\gtrsim 100$ MeV for $10^{-10}\lesssim\epsilon \lesssim
10^{-7}$ \cite{Dent:2012mx}.  Nonthermal DM production necessitates a stronger
gauge coupling, which further suppresses the required kinetic mixing.

Using (\ref{epstriplet}) and (\ref{relic_const}) 
(assuming thermal production of the 
DM), we can eliminate $\epsilon$ and $\alpha_g$ from the predicted
cross section (\ref{sigma_p}) for scattering on protons, 
\beq
\label{direct2}
	\sigma_p = \frac{5.9\times 10^{-43}\text{ cm}^2}
	{F_t(r_2,r_3)^2}\,
	\left(\frac{0.3}{f_3}\right) \left(1+\frac{m_p}{M_\chi}\right)^{-2}
	\left(\frac{M_\chi}{10\text{ GeV}}\right) \left(\frac{100\text{ MeV}}{m_{B_1}}
	\right)^4 ,
\eeq
recalling that the gauge boson masses are related as 
in (\ref{massrels}), depending on the 
choice of Higgs sector.  Scattering from nuclei in direct detection experiments
can either be endothermic $\chi_2N\to \chi_3N$ or exothermic $\chi_3N\to\chi_2N$
events, since both states are populated at present day.  The state
$\chi_1$ does not participate in nuclear scattering.

\begin{figure}[t]\begin{center}
\includegraphics[scale=0.5]{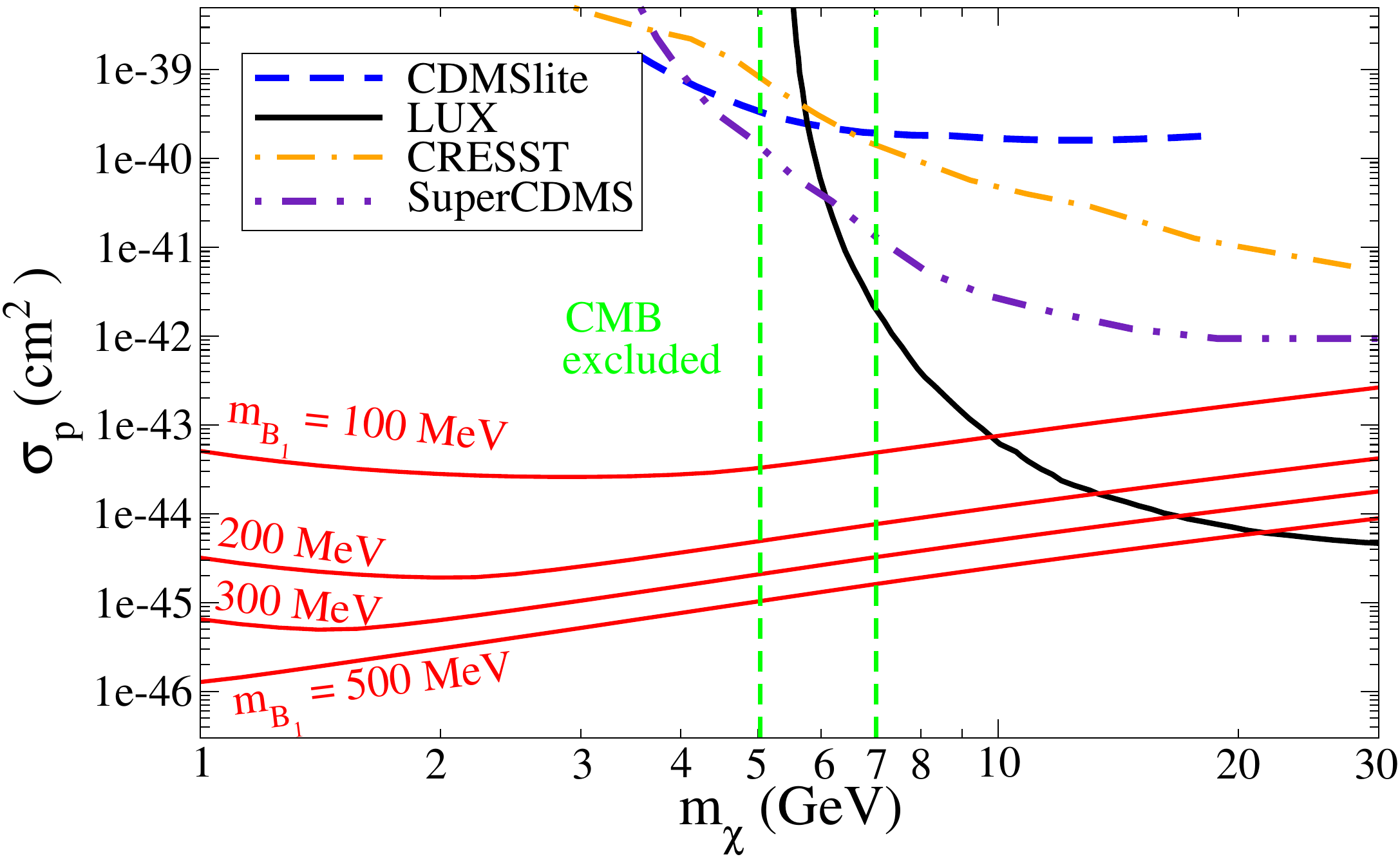} 
\end{center}
\caption{Cross section of triplet DM on protons, 
assuming $\epsilon$ chosen
as in (\ref{epstriplet}) to produce the X-ray line by long-lived excited DM 
decay, as well as the thermal relic value for
$\alpha_g$, eq.\ (\ref{relic_const}).  
Each theoretically predicted curve is labeled by the 
$B_1$ gauge boson mass, 
assuming a dark Higgs sector with two triplets.  As a function
of $m_{B_1}$, the $\sigma_p$ curve reaches a minimum around 
$m_{B_1}=500$ MeV, and increases slightly for larger $m_{B_1}$.
Vertical dashed lines indicate CMB lower limit on $M_\chi$ from
annihilations, depending upon whether $M_{\chi_1}> M_{\chi_2}$
(right line) or $M_{\chi_1}< M_{\chi_2}$ (left line).}
\label{f:dd-triplet1}
\end{figure}

\begin{figure}[t]\begin{center}
\includegraphics[scale=0.5]{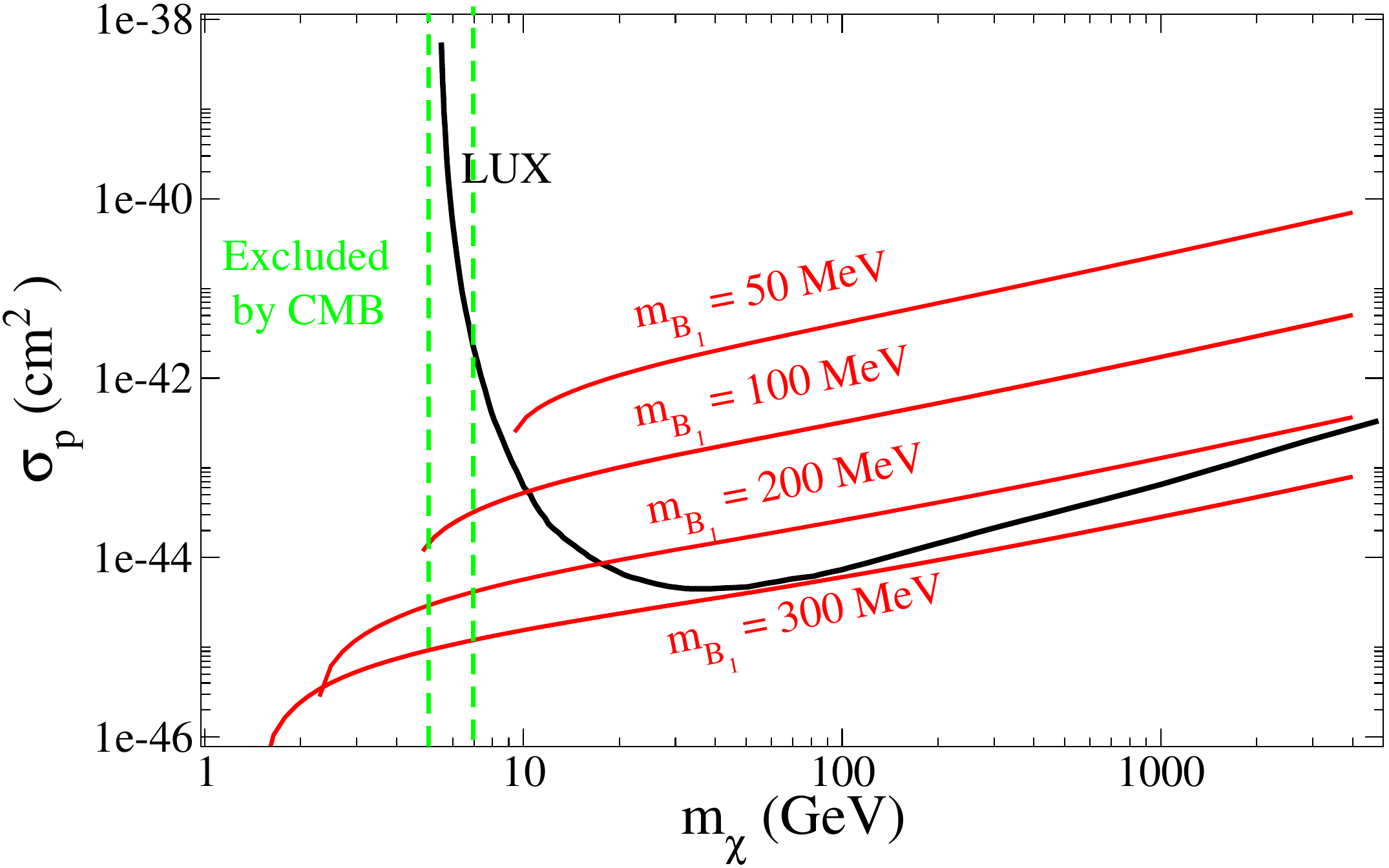} 
\end{center}
\caption{Like figure \ref{f:dd-triplet1} but for a dark Higgs sector of
one triplet and one doublet.}
\label{f:dd-triplet2}
\end{figure}

However, limits on $\sigma_p$ from direct 
detection experiments are expressed in terms of cross sections for 
elastic scattering.  In the approximation that the 
inelasticity modifies
the cross section through the phase space, but not the recoil spectrum, 
the event rate for the combination of endothermic and
exothermic cross sections is equivalent to elastic scattering of the entire
abundance of DM with a cross section on protons of 
\beq\label{directrescale1}
\tilde\sigma_p = \frac{\sigma_p}{\langle v\rangle}\left(
f_2\left\langle \sqrt{v^2-v_\delta^2}\,\Theta(v-v_\delta)\right\rangle+f_3
\left\langle \sqrt{v^2+v_\delta^2}\right\rangle \right)
\eeq
as in eq.\ (\ref{inelastic}).
The rescaled cross section $\tilde\sigma_p$ is shown in figure 
\ref{f:dd-triplet1} for several values of the gauge boson masses for
the two-Higgs-triplet model.   Also shown are the experimental limits, rescaled 
by $(Z/A)^2$ because the DM couples only to charge in our models.
Dark matter masses of $M_\chi\lesssim 10$ GeV are 
allowed for $m_{B_1}=100$ MeV, while larger $M_\chi$
is allowed as $m_{B_1}$ increases. However, this dependence on
$m_{B_1}$ is not
monotonic, due to the factor $F_t(r_2,r_3)^{-2}$ in (\ref{direct2})
and the relations (\ref{massrels}) between the gauge boson masses;
in fact the growth in the allowed value of $M_\chi$ saturates near 
the $m_{B_1}=500$ MeV curve shown, so that higher values than
$M_\chi\sim 20$ GeV are not allowed by the LUX constraint.
For the model with one triplet and one doublet Higgs field
on the other hand, the dependence on $m_{B_1}$ is monotonic,
and for $m_{B_1}>300$ MeV the predicted cross sections fall below
the current LUX limit, as shown in figure 
\ref{f:dd-triplet2}.

Like their doublet DM model counterpart, these
models of decaying triplet DM are potentially constrained by 
the CMB.  Apart from 
the slightly different fractional abundance $f_x$ of excited DM in the
two models, the decay rate is the same, so 
the CMB bounds from decays are robustly satisfied, as in figure \ref{f:cmb-doublet}.  
However, there are related CMB constraints on DM annihilations into SM
particles that apply for the triplet model, since it is a symmetric DM
candidate.  Due to the nonabelian structure of the DM sector, the effective
annihilation cross section does not take the canonical relic density
value in the late
universe.  In appendix \ref{s:cmb} we adapt the limits of 
ref.\ \cite{Madhavacheril:2013cna} to this model and find that 
it must satisfy $M_\chi\gtrsim 5$ GeV
(if $M_{\chi_1}<M_{\chi_2}$) or $M_\chi\gtrsim 7$ GeV ($M_{\chi_2}<M_{\chi_1}$),
depending on the gauge boson masses,  through both the mass splittings and
primordial relative abundances.
These constraints are compatible with limits on the
DM mass from direct detection experiments for gauge boson masses $m_{B_1}\gtrsim
100$ MeV and offer the possibility of detection as CMB limits improve.

\subsection{Triplet XDM}\label{s:tripletxdm}

We next consider the scenario where triplet DM undergoes collisional
excitation followed by fast decays to give the 3.5 keV line.
In this case the kinetic mixing is large enough so that 
the primordial component of the excited DM state $\chi_3$ 
has decayed, and is only repopulated by collisional 
excitation.  As discussed in section \ref{s:doubletxdm}, CMB
constraints required the lifetime to be less than $\sim 10^{12}\,$s.
This requires
\beq\label{eps-tripletxdm}
\epsilon \gtrsim 3.8\times 10^{-9}\,
\frac{(M_\chi/10{\rm\ GeV})}{F_t(r_2,r_3)\,\alpha_g^{3/2}} = (1.9\times 10^{-3}) \left(
\frac{10\text{ GeV}}{M_\chi}\right)^{1/2}\frac{1}{F_t(r_2,r_3)},\eeq
with the latter equality representing the bound for thermal relic DM.

We next consider the rate of upscattering needed to populate $\chi_3$ for 
XDM-like production of the X-ray signal.  Suppose first that the 
entire signal is produced by collisional excitation of a single DM state
$\chi_g$, which could be either $\chi_1$ or $\chi_2$, as both are stable in
the models we consider.
Taking the lower value from equation (\ref{perseussigma}), we find that that
the gauge boson mediating the upscattering $\chi_g\chi_g\to\chi_3\chi_3$ has
mass
\beqa
\label{xdmtripletBmass}
	m_B&\cong& (3\textnormal{ GeV})\left(\frac{\alpha_g f_g}{0.33}\right)^{1/2}
	\left(\frac{v_t\gamma}{1300\textnormal{ km/s}}\right)^{1/4},\\
\label{Bmassthermal}
	&\cong&(37\textnormal{ MeV})\left(\frac{f_g}{0.33}\right)^{1/2}
	\left(\frac{M_\chi}{10\textnormal{ GeV}}\right)^{1/2}
	\left(\frac{v_t\gamma}{1300\textnormal{ km/s}}\right)^{1/4}
\eeqa
where the second line 
assumes the thermal relic value of the gauge coupling. If we consider excitation
of $\chi_2$ DM, then $f_g\sim 2/3$ because of the early universe
decays $\chi_3\to\chi_2\gamma$.  This gives a gauge boson mass 
$m_{B_1}\sim 50$ MeV, very similar to the mass difference needed to account for the
3.55 keV DM mass splitting.  
It is also possible that $\chi_1\chi_1\to\chi_3\chi_3$ scattering produces
the X-ray signal, either as the dominant contribution or along with
$\chi_2\chi_2\to\chi_3\chi_3$ scattering.  In this case, it is
$m_{B_2}$ that is 
$\gtrsim 37$ MeV, and $m_{B_1}$ can be somewhat higher.  It is worth noting that
the mass gap $\delta M_{13}$ may be either larger or smaller than 3.55 keV, 
so the relative contribution of each type of upscattering differs in M31
and the Perseus cluster.  However, these corrections are all of order unity.

Direct detection imposes stringent constraints on these models for DM masses
of order $M_\chi=10$ GeV.  Using (\ref{eps-tripletxdm}) and 
(\ref{xdmtripletBmass}), we find that the cross section for DM-proton inelastic
scattering is 
\beq\label{tripletxdm-dd}
\sigma_p\gtrsim \frac{2.2\times 10^{-47}\textnormal{ cm}^2}{\alpha_g^4 
\,F_t(r_2,r_3)^2} \left(1+\frac{m_p}{M_\chi}\right)^{-2}
\left(\frac{M_\chi}{10\textnormal{ GeV}}\right)^2\left(\frac{0.33}{f_g}\right)^2
\left(\frac{1300\textnormal{ km/s}}{v_t\gamma}\right),\eeq
where $v_t\gamma$ is the appropriate value for the Perseus cluster.
For comparison to elastic scattering experiments, we rescale the cross section
as in equation (\ref{directrescale1}) with $f_3=0$.  
With nonthermal DM production, $\alpha_g$ as small as $0.01$ and somewhat 
lighter DM masses 3-5 GeV avoid current direct detection constraints.

Using the thermal relic value for $\alpha_g$ yields a much stronger constraint
\beq\label{tripletxdm-dd2}
\sigma_p\gtrsim \frac{4\times 10^{-32}\textnormal{ cm}^2}{F_t(r_2,r_3)^2}
\left(1+\frac{m_p}{M_\chi}\right)^{-2}
\left(\frac{10\textnormal{ GeV}}{M_\chi}\right)^2\left(\frac{0.33}{f_g}\right)^2
\left(\frac{1300\textnormal{ km/s}}{v_t\gamma}\right),\eeq
which forces $M_\chi\sim 1$ GeV, below the sensitivity of current direct 
detection experiments.  At these low masses, kinematic suppression of
the scattering rate due to the inelasticity becomes a strong effect and is worth
studying in more detail as more sensitive direct detection experiments come
online.  This cross section can be reduced slightly if $m_{B_1}$ is somewhat
larger than (\ref{Bmassthermal}) but only by about an order of magnitude,
which does not by itself remove this constraint.

However, CMB constraints on DM annihilation will essentially rule out these
models in combination with direct detection constraints.  As described
in appendix \ref{s:cmb}, we find that CMB bounds require 
$M_\chi>6$ GeV for triplet XDM.  This would only be consistent with 
the direct detection constraints on the thermal WIMP model discussed above 
if some combination of parameter choices could drastically reduce the 
direct detection cross section.  Nonthermal DM production with larger
$\alpha_g$ will reduce the DM-proton scattering cross section given in
equation (\ref{tripletxdm-dd}) at the cost of increasing the annihilation
cross section and therefore tightening the CMB constraint.  We would
therefore have to consider replacing the Majorana DM with Dirac DM
and taking an asymmetric DM model in order to make triplet XDM a
viable option.

\section{Discussion}\label{s:discussion}

Nonabelian dark matter models with a light kinetically mixed gauge
boson can naturally incorporate the small mass splitting and coupling
to photons that would be needed for decays of an excited DM state to
explain the 3.5 keV X-ray line.  We find that the DM mass should be
in the approximate range $1-10$ GeV, with the possibility of 
larger masses for decaying triplet DM 
(depending on the dark Higgs content; see fig.\ \ref{f:dd-triplet2}), 
and the gauge bosons masses should
be $\lesssim 1$ GeV.  There are good prospects for direct detection
of the DM by its scattering on protons.  Moreover the kinetic mixing
parameter and gauge boson masses typically fall into regions that will
be probed by the HPS experiment within the next year.

Generally, a given model can exist in either of two regimes,  where
the excited state is metastable and primordial, or else having a
shorter lifetime and produced through inelastic scatterings (XDM
mechanism).  Models of the former kind easily satisfy CMB constraints
from injection of electromagnetic energy during recombination, but the
latter kind  are more strongly constrained by the CMB, needing much
larger values of the kinetic mixing in order to decay well before
recombination. For the doublet XDM model, it requires taking
$M_\chi$ below a few GeV to evade direct detection, while for the
triplet model, since it is a symmetric dark matter candidate, this
loophole is blocked by CMB constraints on annihilations, so that 
the XDM version of the triplet model is ruled out (though asymmetric
Dirac triplet DM could be made acceptable).

In previous literature, nonabelian DM models were explored 
as a means of explaining the anomalous 511 keV gamma ray line from 
the galactic center \cite{Cline:2010kv}.  One could be tempted to 
try to combine
this and the 3.5 keV line in a triplet DM model where there are two
mass splittings corresponding to these energies.  We find (details not
described here) that although it is possible to arrange for the
desired splittings, the relative strengths of the two lines cannot be
correctly reproduced in these models with small multiplets 
because the transition with the
larger energy has too big a rate relative to the smaller one, due to
the larger phase space for the decays.

Similarly, one may wonder whether our model could simultaneously
explain the GeV-scale galactic center excess 
\cite{Hooper:2011ti,Abazajian:2012pn}, since 
general models of DM interacting through kinetically mixed vector
mediators have been shown to give a good fit to the data
\cite{Berlin:2014pya,Cline:2014dwa}.  However the best fit region
is for $m_\chi\sim m_B\sim 30$ GeV, which is generally incompatible
with the parameters we have identified for the X-ray line.  For
example, eq.\ (\ref{eps_doublet}) for the slow-decaying doublet model
implies $\epsilon \sim 10^{-5}/\alpha_g^{3/2}$, giving a cross section
on protons of $\sigma_p \gtrsim 3\times 10^{-44}\,$cm$^2$ that is firmly
excluded by LUX.  More recently it has been pointed out that inclusion
of inverse Compton scattering and brehmsstrahlung contributions can
lead to lower best-fit values $M_\chi\sim 10$ GeV consistent with
leptonic final states \cite{Lacroix:2014eea,Cirelli:2014lwa}.
These authors do not consider the 4-lepton final states that would
arise from pairs of gauge bosons, so we cannot draw any direct
conclusions from their work, but if for example $m_B= 3$ GeV
rather than $\sim 30$ GeV, this would reduce $\epsilon$ by a factor of
100 and $\sigma_p$ by a factor of $10^{4}$, safely below current
direct detection limits.  We leave this interesting question for a
separate study.

The 3.5 keV line awaits confirmation by higher-statistics
observations.  So far the only study to cast doubt on the  observation
is a negative search for the line in our own galactic center
\cite{Riemer-Sorensen:2014yda}; however the conclusions depend upon
uncertain assumptions about the shape of the dark matter halo profile
in this region.  In our study we have pointed out a possible way of
discriminating between and XDM and decaying models for the X-ray line:
since the XDM mechanism depends upon the DM velocity dispersion
through $\langle\sigma v_{\rm rel}\rangle$, one could expect that the
line strength will be relatively stronger from galactic clusters with
higher $v_{\rm rel}$ than from individual galaxies like M31 (see also
\cite{Finkbeiner:2014sja}).  We
quantified this for the predicted signal in fig.\  \ref{f:thermavg}. 
There is already a hint of such an effect in  the present
determinations of the required value of  $\langle\sigma v_{\rm
rel}\rangle$.  It will be interesting to see whether it persists as
the observations improve.
\bigskip

{\bf Acknowledgment.}  We thank Wei Xue for helpful discussions
concerning the galactic center gamma ray excess.  Our work is 
supported by the Natural Sciences and Engineering Research Council
(NSERC) of Canada.

\appendix

\section{Magnetic moments}
\label{app:mu}

We derive the one-loop results for the transition magnetic moments
of the dark matter multiplets, starting with the case of SU(2) triplet
DM.  
In a constant external magnetic field, the triple-gauge interaction
from the kinetic mixing operator takes the form $\epsilon g F_{\mu\nu} B_2^\mu\, B_3^\nu$. 
In the case of equal gauge boson masses,  the effective
operator involving $F_{\mu\nu}$ can be written as 
\beqa
	\epsilon g^3\bar u_2(p)\left[\int {d^{\,4}\ell\over (2\pi)^4}
	{\gamma_\mu(\slashed{\ell}+ M_\chi)\gamma_\nu \over
	(\ell^2+2\ell\cdot p) (\ell^2-m_B^2 + i\epsilon)^2} \right] u_3(p)\,F^{\mu\nu}
\eeqa
setting the momentum of the constant field to zero and ignoring 
the small DM mass splittings. $p$ is the external
fermion momentum which is taken to be on shell, $p^2 = M_\chi^2$.  
The term in brackets becomes
\beqa
	{i\over 16\pi^2}\int_0^1 dx\,(1-x)
	{\gamma_\mu(\slashed{p}(1-x)+ M_\chi)\gamma_\nu \over 
	x^2\, M_\chi^2 + (1-x)\mu^2}
\label{paramint}
\eeqa
after doing the momentum integral.  By anticommuting half of the $\slashed{p}$
term through each gamma matrix and using the Dirac equation, we find that 
$\slashed{p}\to -m$ plus terms that are symmetric under $\mu\leftrightarrow\nu$,
hence vanish under contraction with the field strength.  The $x$ integral can be
done, resulting in the transition magnetic moment
\beqa	
	\mu_\times = {\epsilon g^3\over 16\pi^2\, M_\chi} 
	F_t(m_B^2/M_\chi^2)
\eeqa
where
\beqa
	F_t(r) = {1-r\over 2}\ln{1\over r} -1 + 
	{3-r\over \sqrt{4/r-1}}\, 
	\tan^{-1}\sqrt{4/r-1} 
\label{ftfun0}
\eeqa
It has leading behavior $\frac12\ln(1/r)$ at small $r$.
For the case of two different gauge boson masses in the loop,
we define $r_i = m_{B_i}^2/M_\chi^2$ and obtain in place of
(\ref{paramint})
\beqa
	{i\over 16\pi^2}\int_0^1 dx\,\int_0^{1-x}dy\,
	{\gamma_\mu(\slashed{p}(1-x)+ M_\chi)\gamma_\nu \over 
	x^2\, M_\chi^2 + y m_{B_1}^2 + (1-x-y)m_{B_2}^2}
\label{paramint2}
\eeqa
leading to 
\beqa
	F_t(r_1,r_2) &=& 
{\sqrt{(4-r_1)}\,r_1^{3/2}\tan^{-1}\sqrt{4/r_1-1}-
\sqrt{(4-r_2)}\,r_2^{3/2}\tan^{-1}\sqrt{4/r_2-1}\over 2(r_1-r_2)}
\nonumber\\
&+&{(r_1^2-2 r_1) \ln r_1  - (r_2^2-2 r_2) \ln r_2
\over 4(r_1 -  r_2)}-\frac12
\label{ftfun}
\eeqa
One can show that (\ref{ftfun}) reduces to (\ref{ftfun0}) in the 
limit $r_1\to r_2 = r$.

The above results can be generalized to other DM representations of
SU(N) by including the group theory factor
\beq
	G = f^{\hat a bc} T^c_{2i} T^b_{i1}
\label{gtfact}
\eeq
assuming that the DM mass eigenstates 
labeled $1,2$ correspond to the generators $T^b_{i1}$, $T^c_{2i}$
respectively, and $\hat a$ denotes the gauge boson that kinetically
mixes with the photon.  One must sum over gauge bosons with masses $m_{B_b}$
and $m_{B_c}$ as well as the internal DM state with mass $M_{\chi_i}$.
If the mass differences can be neglected in the loop integral (as fig.\ \ref{f:moment} shows is
often a good approximation) then the sums over $i,b,c$ can be done
directly in (\ref{gtfact}), giving
\beq
	G = \frac i2 C_2(A) T^{\hat a}_{21}
\label{gtfact2}
\eeq
where $C(A)$ is the quadratic Casimir invariant of the adjoint
representation.  The multiplicative correction factor  (\ref{gtfact2}) is unity for the triplet model in SU(2).

In the SU(2) doublet DM model, the transition magnetic moment gets
nearly canceling contributions from both $\chi_1$ and $\chi_2$ in the loop, 
such that the result is suppressed by $\delta M_\chi$.  One must
therefore be more careful in distinguishing the incoming and outgoing
fermion momenta $p_{1,2}$, and keeping the dependence on the
photon momentum $q = p_1-p_2$.  The induced operator is
\beqa
	&&\epsilon g^3\bar u_1(p_1)\left[\int {d^{\,4}\ell\over (2\pi)^4}
	{\gamma_\mu(\slashed{\ell}+\slashed{p}+ m_1)\gamma_\nu \over
	((\ell+p)^2-m_1^2) ((\ell-q/2)^2+m_B^2)((\ell+q/2)^2+m_B^2)} 
	\right] u_2(p_2)\,F^{\mu\nu}(q)\nonumber\\
	&-& \left\{ m_1\to m_2\right\}
\eeqa
where $p=\frac12(p_1+p_2)$.  Introducing Feynman parameters $x$ and
$y$ for the two gauge boson propagators (hence $(1-x-y)$ for the
fermion) we find the result
\beq
	\mu_\times = {\epsilon g^3\over 16\pi^2}
	{\delta M_\chi\over M_\chi^2}\, F_d(r)
\eeq
to leading order in $\delta M_\chi$
for the transition magnetic moment, with $r=m_B^2/M_\chi^2$ and
\beqa
	F_d(r) &=& \int_0^1 dx \int_0^{1-x} dy\, {(1-x-y) +r(x+y)\over
	\left((1-x-y)^2 + r(x+y)\right)^2}\nonumber\\
	&=& {2-r\over r(4-r)} + 2{\tan^{-1}\sqrt{4/r-1}\over
	\sqrt{r}(4-r)^{3/2}}
\label{fdfun}
\eeqa
which has leading behavior $1/(2r)$ at small $r$.  This implies
that for $m_B\ll M_\chi$, the transition moment in the doublet model
goes like $\delta M_\chi/m_B^2$ instead of $1/M_\chi$.

\section{CMB bounds from DM annihilation}\label{s:cmb}

In this appendix, we give details concerning 
the constraints on the mass $M_\chi$ of triplet DM models described
in section \ref{s:triplet} coming from the cosmic microwave background.
Combined limits from
Planck, WMAP9, ACT, and SPT (plus low-redshift data) constrain DM with 
canonical annihilation cross section $\sigma v=3\times 10^{-26}$ cm$^3/$s to have
mass $M_\chi>(26$ GeV$)f$, where $f$ is the effective energy deposition efficiency
\cite{Madhavacheril:2013cna}. For ``XDM-like'' annihilation processes
$\chi\chi\to BB$ followed by $B_1\to e^+e^-$, the efficiency is $f=0.67$ for 
DM masses near 10 GeV \cite{Madhavacheril:2013cna}.  This efficiency is 
larger than for $B_1\to 2\mu$ or $B_1\to 2\pi$ decay channels 
when they are allowed.
Since we are mostly concerned with lighter gauge boson masses, we 
conservatively take $f=0.67$.  However, the constraints we find below 
are loosened somewhat for $m_{B_1}>2m_\mu$.

We account for several additional
effects in our models.  First, as described in section \ref{s:nonabelian},
we use the adjusted value of the thermal relic cross section, which is
approximately 0.7 of the canonical value at DM mass of 10 GeV but increases
to the canonical value at 5 GeV DM mass.  

There are other effects due to the nonabelian structure of the dark sector.
The energy deposition efficiency
is reduced by the fact that the relic abundance is determined by
the total annihilation cross section for $\chi\chi\to BB$, but only $B_1$
gauge bosons decay into SM particles.  Therefore, we count only annihilations
with $B_1$ final particles toward the energy deposition efficiency (weighting
final states with a single $B_1$ with 1/2). Finally, the relative abundances of
the DM species differ in the present universe relative to those
at chemical freeze out due
to kinetic equilibration at lower temperatures during the Big Bang.
As a result of these two effects, the effective average annihilation cross
section is modified in the late universe compared to the canonical value.
As given in the appendix of \cite{Cline:2010kv}, the color- and spin-averaged
and summed square amplitude for the total annihilation cross section 
(assuming equal abundances for each ``color,'' or DM state) is 
$|\mathcal M_{tot}|^2=25g^4/6$, not including any overcounting or 
symmetry factors for identical initial or final particles.  This is composed
of amplitudes for two types of processes, $\mathcal M_{12\to 12}$, which 
involves $t$- and $s$-channel diagrams, and $\mathcal M_{22\to 11}$, with 
$t$- and $u$-channel parts.  As we approximate vanishing gauge boson masses,
the amplitudes are equal when gauge indices are permuted.
The effective squared amplitude at late times, including all permutations
and weighting $\mathcal M_{12\to 12}$ contributions by 1/2, is
\beqa
|\mathcal M_{eff}|^2 &=& \frac 42 f_1f_2|\mathcal M_{12\to 12}|^2 +
\frac 42 f_1f_3|\mathcal M_{13\to 13}|^2+f_2^2 |\mathcal M_{22\to 11}|^2
+f_3^2 |\mathcal M_{33\to 11}|^2\\
&=& \left[\frac 92 f_1(1-f_1)+4(f_2^2+f_3^2)\right] g^4.\label{Meff}
\eeqa
We therefore rescale the energy deposition efficiency by 
$|\mathcal M_{eff}|^2/|\mathcal M_{tot}|^2$.

In the decaying triplet model, the fractional relic abundances $f_a$ are
set by kinetic freezeout in the Big Bang.  For fiducial values 
$f_1=f_2=f_3=1/3$, we find the bound $M_\chi\gtrsim 6$ GeV.  For specific values
of the mass splittings and gauge boson masses, the relative abundances change
somewhat; representative abundances are $f_1=0.66$, $f_2=0.23$, $f_3=0.11$
if $M_{\chi_1}<M_{\chi_2}$ or $f_1=0.23$, $f_2=0.66$, $f_3=0.11$ if 
$M_{\chi_2}<M_{\chi_1}$.  These lead to CMB bounds of 5 GeV and 7 GeV respectively.

In the XDM model, the primordial $\chi_3$ population decays early, adding to 
the $\chi_2$ abundance.  Therefore, the three cases described above 
have $f_1=1/3$, $f_2=2/3$, $f_3=0$ with $M_\chi>8$ GeV,  $f_1=0.66$, $f_2=0.34$, 
$f_3=0$ with $M_\chi>6$ GeV, and $f_1=0.23$, $f_2=0.77$, $f_3=0$ with
$M_\chi>9$ GeV.  These values are inconsistent with the direct detection bounds
for the XDM model.

\bibliographystyle{JHEP}
\bibliography{xraydecays}

\end{document}